\documentclass[prf,longbibliography,onecolumn,superscriptaddress,amsmath,amssymb,aps,floatfix]{revtex4-2}

\usepackage{graphicx}% Include figure files
\usepackage{dcolumn}% Align table columns on decimal point
\usepackage{bm}% bold math
\usepackage{xcolor}

\DeclareMathOperator{\erfc}{erfc}
\begin{document}
%\preprint{APS/123-QED}
\def\anu{\textcolor{magenta}}
\def\check{\textcolor{blue}}
\title{\textbf{Trapping and Transport of Inertial Particles in a Taylor-Green Vortex: Effects of Added Mass and History Force} 
}% 

\author{Prabhash Kumar}
\affiliation{Department of Applied Mechanics \& Biomedical Engineering, Indian Institute of Technology Madras, Chennai 600036, India}%Lines break automatically or can be forced with \\
\author{Anu V. S. Nath}%
\affiliation{Department of Applied Mechanics \& Biomedical Engineering, Indian Institute of Technology Madras, Chennai 600036, India}

\author{Mahesh Panchagnula}
\affiliation{Department of Applied Mechanics \& Biomedical Engineering, Indian Institute of Technology Madras, Chennai 600036, India}
% \affiliation{
%  second institution for this author
% }%
\author{Anubhab Roy}
\email{Contact author: anubhab@iitm.ac.in}
\affiliation{Department of Applied Mechanics \& Biomedical Engineering, Indian Institute of Technology Madras, Chennai 600036, India}


\begin{abstract}
We investigate the dynamics of \textit{small} inertial particles in a two-dimensional, steady Taylor-Green vortex flow. A classic study by Taylor (2022) showed that heavy inertial point particles (having density parameter $R = 1$) are trapped by the flow separatrices when the particle Stokes number $\textrm{St}$, which measures the particle's inertia, is less than $1/4$. Here, we consider finitely dense particles, incorporating the previously neglected effects of added mass and the Boussinesq-Basset history force. Using linear stability analysis near stagnation points, we determine the critical parametric conditions in the $\textrm{St}$--$\textrm{R}$ plane that leads to particle trapping within vortex cells. We identify additional stagnation points perceived by inertial particles, beyond the traditional ones at vortex cell corners, when the added mass effect is included, and we analyze their stability. Numerical analysis of the full nonlinear system confirms the existence of distinct particle behaviours--trapped, diffusive, and ballistic--depending on initial conditions, consistent with Nath \textit{et al.} (2024), with modifications due to added mass effect. We delineate the regions in the $\textrm{St}$--$\textrm{R}$ plane where these behaviours dominate based on the prominent particle dynamics. However, when both the history force and added mass effect are included, all particles exhibit ballistic motion regardless of $\textrm{St}$ and $\textrm{R}$.
\end{abstract}

\maketitle
%##########################################
\section{\label{sec:Introduction}Introduction}
The transport and dispersion of \textit{small} inertial particles in fluid flow play a crucial role in numerous physiological, environmental, and engineering applications. Examples include particulate transport in pulmonary airways \cite{kleinstreuer2010airflow}, blood flow in arteries \cite{martel2014inertial}, the formation of cloud droplets in turbulent atmospheres \cite{falkovich2002acceleration}, and the movement of dust and debris in hurricanes \cite{sapsis2009inertial}. Additionally, inertial particle transport is fundamental to various industrial processes such as spray drying, pollution control, and slurry transport, among others \cite{michaelides2022multiphase}.

\citet{maxey1983equation} derived an equation of motion to describe the dynamics of a small, rigid sphere in a nonuniform flow as
\begin{align}
 m_{p}\,\frac{d\mathbf{v}}{dt} &=  m_{f}\,{\frac{D\mathbf{u}}{Dt}} + \frac{m_f}{2}\,\left(\frac{D\mathbf{u}}{Dt}-\frac{d\mathbf{v}}{dt}\right) - 6\,\pi\,\mu \, a\,\left(\mathbf{v}-\mathbf{u}\right) \nonumber\\
 &- 6\,a^2\,\rho_f\,\sqrt{\pi\,\nu}\,\left[\int_{t_0}^{t}\left(\frac{d\mathbf{v}}{dt'}-\frac{d\mathbf{u}}{dt'}\right)\,\frac{dt'}{\sqrt{t-t'}}+\frac{\mathbf{v}_{0}-\mathbf{u}_{0}}{\sqrt{t-t_0}}\right]~,
 \label{Maxey_Riley equation}
 \end{align}
 where $m_p$ and $m_f$ are the particle mass and the mass of the displaced fluid, respectively. The densities of the particle and fluid are denoted by $\rho_p$ and $\rho_f$, respectively. The particle's Lagrangian velocity is represented by $\mathbf{v}$, while $\mathbf{u}$ is the fluid velocity sampled at the particle's position $\mathbf{x}$, where $d\mathbf{x}/dt = \mathbf{v}$. The particle has a radius $a$, and the fluid properties are characterised by the dynamic viscosity $\mu$ and the kinematic viscosity $\nu$. The initial time is denoted by $t_0$, with the corresponding initial velocities of the particle and fluid given by $\mathbf{v}(t=t_0) = \mathbf{v}_0$ and $\mathbf{u}(t=t_0) = \mathbf{u}_0$, respectively. The derivatives: $ D/{Dt}=\partial/\partial t+\mathbf{u}\bm{\cdot \nabla}$, represents the total derivative following a fluid element, and $d/{dt}=\partial /{\partial t}+\mathbf{v}\bm{\cdot \nabla}$, denotes the total derivative following the particle trajectory. The first two terms on the right-hand side of equation~(\ref{Maxey_Riley equation}) correspond to inertial forces, which include the pressure gradient force and the added mass effect. The remaining two terms are viscous forces, comprising Stokes drag and the Boussinesq-Basset history force (or simply `history force'). In its original form, equation~(\ref{Maxey_Riley equation}) includes Faxén correction terms as well, which are neglected here as they become insignificant for particles significantly smaller than the flow length scale. We also neglect the gravitational effects in the current analysis.
  
The presence of the history force makes the Maxey-Riley equation an integro-differential equation, rendering its analytical solution highly non-trivial. Additionally, numerical solutions are computationally expensive and require significant memory resources~\cite{daitche2013advection,prasath2019accurate}. To avoid these challenges, two simplified models of the Maxey-Riley equation are widely adopted in the literature. One of these models is given as
 \begin{equation}\label{Simplified_Maxey-Riley_equation}
     m_{p}\frac{d\mathbf{v}}{dt} = -6\,\pi\,\mu\, a\,\left(\mathbf{v}-\mathbf{u}\right), 
 \end{equation}
which applies in the limit of heavy particles, where the particle-to-fluid density ratio is very large ($\rho_p \gg \rho_f$). This model is commonly used to study the transport of particles and droplets in various atmospheric turbulent flows~\cite{wilkinson2005caustics,bec2003fractal,bec2005clustering}. The other model is a reduced-order Maxey-Riley equation, originally derived by \citet{ferry2001fast}, also known as the inertial equation~\cite{haller2008inertial}, given by
\begin{equation}
    \mathbf{v} = \mathbf{u} + \tau_{p}\,\left(\frac{\rho_f}{\rho_p}-1\right)\frac{D\mathbf{u}}{Dt} +\mathcal{O}(\tau_{p}^{3/2}),
\end{equation}
where $\tau_{p}=m_{p}/(6\,\pi\,\mu\, a)$ represents the particle relaxation time, characterizing the time scale over which the particle adjusts its velocity to the surrounding fluid motion. The inertial equation is valid in the limit $\tau_p \ll L_0/U_0$, implying that the particle velocity closely follows the flow velocity with an inertial correction due to particle inertia. This equation is derived through asymptotic analysis under the condition $\tau_p \ll L_0/U_0$, where the influence of Stokes drag and inertial forces occur at $\mathcal{O}(\tau_p)$ while the influence of the history force appears only at higher-order terms. The original derivation by \citet{ferry2001fast} suggested that the correction term would be of the order $\mathcal{O}(\tau_p^{3/2})$ when accounting for the history force. However, in the absence of the history force, the correction terms can be weaker, of the order $\mathcal{O}(\tau_p^{2})$~\cite{haller2008inertial}.

Several researchers have used various forms of the Maxey-Riley equation to investigate the dynamics of inertial particles. Here, we summarize key studies focused on the particle dynamics in vortical and rotating flows. \citet{raju1997dynamics} investigated the motion of small, spherical particles in various basic flow patterns like vortical and stagnation point flows. Using the Maxey-Riley equation (excluding history effects), they analyzed particle dynamics over a wide range of particle-to-fluid density ratios and Stokes numbers, \textrm{St} (ratio of particle relaxation time scale and the flow time scale). The study examined how density ratio, Stokes number, and gravity influence particle ejection, entrapment, and accumulation of particles in these basic flow structures. Through both analytical and numerical approaches, the authors demonstrated that optimal ejection and entrapment typically occur at intermediate Stokes numbers. \citet{crisanti1990passive} examined the transport of particles in a two-dimensional (2D), steady Taylor-Green (TG) vortex flow. Their findings showed that buoyant particles ($\rho_p < \rho_f$) move toward the vortex centre, while dense particles ($\rho_p > \rho_f$) exhibit chaotic trajectories leading to long-term diffusion, referred to as ``inertial diffusion''. They compared this inertial diffusion with the diffusion caused by thermal fluctuations in the fluid, observing similar scaling behaviours in both cases despite the distinct underlying mechanisms. The study highlights that even small density differences between the particle and fluid can lead to significant changes in particle behaviour. Similarly,~\citet{crisanti1992dynamics} examined the transport characteristics of inertial particles in anisotropic as well as time-periodic cellular flows, noting that dense particles remain chaotic and diffusive. However, buoyant particles can exhibit either chaotic or periodic behaviour depending on the flow parameters. The authors identified a scaling relationship between diffusion coefficients and density differences, drawing parallels to systems influenced by additive noise that model thermal fluctuations. Additionally, \citet{wang1992chaotic} explored the chaotic dynamics of inertial particles in both steady cellular flow (i.e., the TG vortex) and a time-evolving shear mixing layer. Their findings indicate that particle motion can be chaotic even in steady, laminar flows, leading to dispersion patterns similar to those observed in turbulent flows. The study explores the relationship between chaotic motion—characterized by fractal dimensions and Lyapunov exponents—and the rate of particle dispersion and mixing efficiency. The findings reveal a complex interplay where higher dispersion rates correspond to lower mixing efficiency. \citet{nath2022transport} investigated the dynamics of heavy inertial particles ($\rho_p \gg \rho_f$) in a TG vortex flow, modelling their motion using only the Stokes drag term in the Maxey-Riley equation (see equation~\ref{Simplified_Maxey-Riley_equation}). Their study confirmed earlier findings~\cite{taylor1940notes} that particles with $\textrm{St} < 1/4$ become trapped by the vortex separatrices. Recently, \citet{nath2024irregular} revisited this problem, extending the work of \citet{wang1992chaotic} while still focusing on heavy particles. They uncovered unexpected behaviour in the TG vortex flow, demonstrating that the system is non-ergodic as the particles can exhibit trapped, diffusive, or ballistic dispersion depending on their initial conditions. They draw an analogy between the dynamics of inertial particles in a TG vortex flow and those of a soft Lorentz gas \citep{klages2019normal}, where inertial particles are ``scattered'' by vortex cells as they traverse stagnation regions. Furthermore, the study classified particle trajectories into two distinct types—regular or chaotic—further reinforcing the non-ergodic nature of the system. Their findings revealed that particle dispersion and effective diffusivity exhibit irregular, non-monotonic, and sometimes discontinuous variations with the Stokes number, challenging previous results. In contrast to \citet{wang1992chaotic}, they observed chaotic motion around $\textrm{St} = \mathcal{O}(1)$  for heavy inertial particles. 
A recent study by \citet{dagan2025analytical} derived novel analytical solutions for the dispersion of heavy inertial particles in the TG vortex flow. By assuming low Stokes numbers, the authors simplified the governing equations and obtained explicit mathematical expressions for particle trajectories and velocities. Their analysis considered both freely moving particles and those influenced by external forces, such as magnetic fields. The analytical predictions were also validated against numerical simulations. Note that, all these previous studies have neglected the effects of the history force due to the challenges associated with its incorporation.

\citet{lasheras1994dynamics} conducted an asymptotic perturbation analysis in the limit of small Stokes number $(\textrm{St}\ll 1)$, considering the influence of hydrodynamic forces, including the history force, on particle dynamics in an isolated Rankine vortex. The particle inertia is characterized by the non-dimensional Stokes number $\textrm{St} = \tau_p\, U_0/L_0$. They demonstrated that history force appears as a higher-order term in the analysis and can be ignored in the limiting scenario ($\textrm{St} \ll 1$). In the Rankine vortex flow, the study revealed that a particle's trajectory is strongly influenced by its density relative to the fluid. Dense particles tend to move away from the vortex centre. In contrast, buoyant particles may become trapped near the centre, forming a stable equilibrium point. \citet{druzhinin1994influence} investigated the motion of small spherical particles in 2D model flows, highlighting the often-ignored role of the history force. They derived an approximate analytical solution for particle velocity, incorporating inertial effects and the history force, for both an axisymmetric vortex with uniform vorticity (circular vortex) and a cellular TG vortex flow. Numerical simulations validated their analytical solutions. For the circular vortex, they found that particles follow a spiral trajectory due to inertia, resulting in a slow radial drift across streamlines, consistent with previous studies. Buoyant particles drift toward the vortex centre, while dense particles move outward. However, the history force altered both the rotation and drift rates, with its effect on drift being more pronounced than on rotation. In the cellular flow, the study highlighted the strong impact of the history force on particle trajectories, particularly its role in separatrix crossings and unbounded motion of dense particles. Through experimental and theoretical analyses, \citet{candelier2004effect} highlighted the significant role of the history force in the radial migration of a small, heavy particle in a rotating fluid. They demonstrated that the history force strongly influences the particle's ejection rate from the vortex centre, a result confirmed experimentally, underscoring its importance, particularly when the particle's density is close to that of the fluid. The study also highlights the importance of using the correct form of the history force, emphasizing the time derivative of the fluid velocity following the particle's path rather than the acceleration of the fluid at the particle's location, even at low Reynolds numbers.\citet{daitche2015role} investigated the influence of the history force on inertial particles in turbulent flows, demonstrating its impact on key aspects such as slip velocity, acceleration, preferential concentration (particle clustering), and collision rates. They found that the history force causes particles to stay closer to the fluid velocity, making them behave more like tracers and thereby reducing clustering in the flow. However, they noted that this effect becomes negligible when the particle is much heavier than the fluid, particularly for particle-to-fluid density ratios of $ \mathcal{O}(10^{3}) $.

This study investigates the dynamics of inertial particles in the TG vortex by modelling their motion using equation~(\ref{Maxey_Riley equation}). The primary objective is to explore the behaviour of dense particles over a broader range of particle-fluid density ratios and to examine the influence of inertial forces and the often-neglected history force on their dynamics. This article is structured as follows: In section~\ref{sec:ProblemFormulation}, we discuss the flow field, outline our objectives, and list the assumptions we made while also examining the non-dimensional form of the equations of motion along with relevant non-dimensional numbers. Before we dive into the analysis of particle dynamics in the TG vortex, we first explore their behaviour near stagnation points—the fundamental components of the TG vortex—using linear stability analysis in section~\ref{sec:ParticleDynamicsNearStagnantPoint}. Next, in section~\ref{sec:Particle dynamics in TG vortex flow}, we linearize the full equation of motion around the fixed points of the TG vortex to investigate the leading-order dynamics. We also discuss the emergence of additional fixed points due to added mass effects in section~\ref{sec:additional stagnant points}. In \S~\ref{sec5}, we conduct a numerical analysis to study particle dynamics within the fully nonlinear TG vortex flow in the absence of history force and analyse the effects of history force on the dynamics in \S~\ref{sec:NumeicalResults}. Finally, we highlight the key findings and provide concluding remarks in section~\ref{sec:Conclusion}.
%%%%%%%%%%%%%%%%%%%%%%%%%%%%%%%%%%%%%%%%%%%%%%%%%%%%%%%%%%%%%%%%%%%%%%%%%%%%%%%%%%%%%%%%%%%%%%%%%
\section{\label{sec:ProblemFormulation}Problem Formulation}
The TG vortex represents a 2D, steady, laminar, and incompressible flow field, characterized by the stream function
\begin{equation}\label{Eq:TG_streamfunction}
    \psi^*\left(\textrm{x}^*,\;\textrm{y}^*\right) = U\,L\,\sin{\left(\frac{\textrm{x}^*}{L}\right)}\,\sin{\left(\frac{\textrm{y}^*}{L}\right)}.
\end{equation}
Here, $(\, )^*$ denotes dimensional quantities. The TG flows features doubly periodic arrays of cells comprising four counter-rotating vortices of cell size $L$, arranged in an unbounded 2D space. The points where the flow velocity vanishes are known as fixed points. For the TG vortex we have considered, these fixed points are located at ($n\, \pi, m\, \pi$), and ($n\, \pi+\pi/2, m\, \pi+\pi/2$) for integers $n$ and $m$. The maximum velocity, $U$, occurs along the boundaries of the vortices. Each vortex cell features fixed points at its corners and center, while each cell boundaries connecting the corner fixed points are flow separatrices. Additionally, the flow exhibits periodic repetition in both the \textrm{x} and \textrm{y} directions. Fluid elements orbit around the streamlines with regular periodic motion. The period of this motion depends on the initial positions of the fluid elements: it is $2\pi \,L/U$ near the vortex center and becomes infinite near the cell boundaries.

\begin{figure}[bth]
    \centering
    \includegraphics[width=0.8\linewidth]{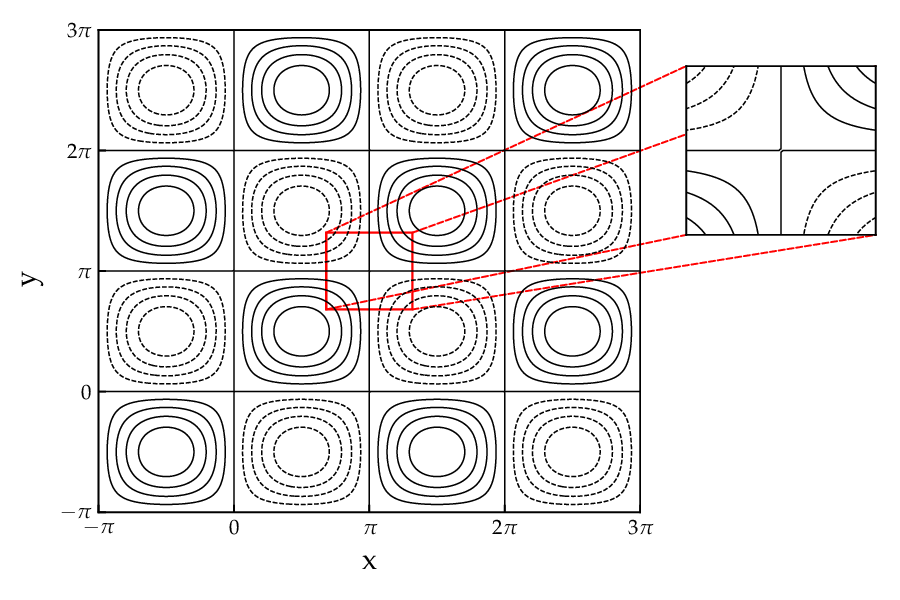}
    \caption{The TG vortex flow displays clear streamlines. The inset provides a zoomed view of the flow patterns near a stagnant point. Solid contour lines indicate anti-clockwise rotation, while dashed contour lines represent clockwise rotation.}
    \label{fig:Streamlines_TG_Flows}
\end{figure}

We use the characteristic flow length scale $L$ and velocity scale $U$ to non-dimensionalise the system. The velocity components for the TG vortex are thus given in non-dimensional as $\mathrm{u}_\mathrm{x}=\sin{\mathrm{x}}\,\cos{\mathrm{y}}$ and $\mathrm{u}_\mathrm{y}=-\cos{\mathrm{x}}\,\sin{\mathrm{y}}$. Here, variables without $(\, )^*$ represent non-dimensional quantities. In FIG.~\ref{fig:Streamlines_TG_Flows}, we illustrate the streamlines of TG flow in the $[-\pi,3\pi] \times [-\pi,3\pi]$ non-dimensional space. Fluid parcels slide along the flow separatrices as they approach the corner fixed points. At the same time, dense inertial particles ($\rho_p > \rho_f$) may cross these fixed points or separatrices depending on some parametric conditions.

The objective of this study is to investigate the dynamics of denser ($\rho_p > \rho_f$) inertial particles in the TG vortex flow described by equation~(\ref{Eq:TG_streamfunction}). We model the inertial particles as \textit{small}, rigid spheres with a radius $ a \ll L $, and we assume that the particle suspension is dilute enough so that particles do not interact with one another and the feedback effect from the particles to the fluid is minimal. This allows us to treat the system as one-way coupled. Under this assumption, we can utilize the Maxey-Riley equation~(\ref{Maxey_Riley equation}) to model the particle motion. After rearranging the terms in equation~(\ref{Maxey_Riley equation}) and substituting the velocity field from equation~(\ref{Eq:TG_streamfunction}), along with scaling by the characteristic flow scales $ U_0 $ and $ L_0 $, the governing equations for particle dynamics can be expressed in a non-dimensional, component form as
% By defining $ U_0 $ as the characteristic flow scale and $ L_0 $ as the characteristic length scale, we can rewrite equation~(\ref{Maxey_Riley equation}) in a non-dimensional form as follows:
% \begin{widetext}
%     \begin{subequations}
%     \begin{align}
%     \frac{d\mathrm{v}_\mathrm{x}}{dt} = & -\frac{\mathrm{R}}{\mathrm{St}} \left\{\mathrm{v}_\mathrm{x} - \sin{\mathrm{x}} \left[\frac{3(1-\mathrm{R})}{\mathrm{R}}\mathrm{St}\cos{\mathrm{x}}+\cos{\mathrm{y}}\right]\right\} \nonumber \\
%     & - \frac{\kappa}{\sqrt{\pi}} \left[\frac{\mathrm{v}_{x_{0}} - \sin{\mathrm{x}_0}\cos{\mathrm{y}_0}}{\sqrt{t}}+\int_{0}^{t} \frac{1}{\sqrt{t-t'}}\left(\frac{d\mathrm{v}_\mathrm{x}}{dt'} - \mathrm{v}_\mathrm{x}\cos{\mathrm{x}}\cos{\mathrm{y}} + \mathrm{v}_\mathrm{y}\sin{\mathrm{x}}\sin{\mathrm{y}}\right)dt' \right],\label{Eq:Equation_of_motion_TG_Flows_X}\\ 
%     \frac{d\mathrm{v}_\mathrm{y}}{dt} = & -\frac{\mathrm{R}}{\mathrm{St}} \left\{\mathrm{v}_\mathrm{y} - \sin{\mathrm{y}} \left[\frac{3(1-\mathrm{R})}{\mathrm{R}}\mathrm{St}\cos{\mathrm{y}}-\cos{\mathrm{x}}\right]\right\} \nonumber \\
%     & - \frac{\kappa}{\sqrt{\pi}} \left[\frac{\mathrm{v}_{y_{0}} + \cos{\mathrm{x}_0}\sin{\mathrm{y}_0}}{\sqrt{t}}+\int_{0}^{t} \frac{1}{\sqrt{t-t'}}\left(\frac{d\mathrm{v}_\mathrm{y}}{dt'} - \mathrm{v}_\mathrm{x}\sin{\mathrm{x}}\sin{\mathrm{y}} + \mathrm{v}_\mathrm{y}\cos{\mathrm{x}}\cos{\mathrm{y}}\right)dt' \right],\label{Eq:Equation_of_motion_TG_Flows_Y}
% \end{align}
% \label{Eq:Equation_of_motion_TG_Flows}
% \end{subequations}
% \end{widetext}
\begin{widetext}
    \begin{subequations}
    \begin{align}
    \frac{d\mathrm{v}_\mathrm{x}}{dt} = & -\frac{\mathrm{R}}{\mathrm{St}}\, \left(\mathrm{v}_\mathrm{x}-\sin{\mathrm{x}}\, \cos{\mathrm{y}} \right)+\frac{3\, (1-\mathrm{R})}{2}\, \sin{2\,\mathrm{x}}\nonumber \\
    & - \frac{\kappa}{\sqrt{\pi}} \left[\frac{\mathrm{v}_{x_{0}} - \sin{\mathrm{x}_0}\cos{\mathrm{y}_0}}{\sqrt{t}}+\int_{0}^{t} \frac{1}{\sqrt{t-t'}}\left(\frac{d\mathrm{v}_\mathrm{x}}{dt'} - \mathrm{v}_\mathrm{x}\cos{\mathrm{x}}\cos{\mathrm{y}} + \mathrm{v}_\mathrm{y}\sin{\mathrm{x}}\sin{\mathrm{y}}\right)dt' \right],\label{Eq:Equation_of_motion_TG_Flows_X}\\ 
    \frac{d\mathrm{v}_\mathrm{y}}{dt} = & -\frac{\mathrm{R}}{\mathrm{St}}\, \left(\mathrm{v}_\mathrm{y}+\sin{\mathrm{y}}\, \cos{\mathrm{x}} \right)+\frac{3\, (1-\mathrm{R})}{2}\, \sin{2\,\mathrm{y}}\nonumber \\
    & - \frac{\kappa}{\sqrt{\pi}} \left[\frac{\mathrm{v}_{y_{0}} + \cos{\mathrm{x}_0}\sin{\mathrm{y}_0}}{\sqrt{t}}+\int_{0}^{t} \frac{1}{\sqrt{t-t'}}\left(\frac{d\mathrm{v}_\mathrm{y}}{dt'} - \mathrm{v}_\mathrm{x}\sin{\mathrm{x}}\sin{\mathrm{y}} + \mathrm{v}_\mathrm{y}\cos{\mathrm{x}}\cos{\mathrm{y}}\right)dt' \right],\label{Eq:Equation_of_motion_TG_Flows_Y}
\end{align}
\label{Eq:Equation_of_motion_TG_Flows}
\end{subequations}
\end{widetext}
with $\mathrm{v}_\mathrm{x}=d\mathrm{x}/dt$ and $\mathrm{v}_\mathrm{y}=d\mathrm{y}/dt$. Here the initial time is set to zero, i.e., $t_0=0$, and the corresponding variables are notated as $\mathrm{x}_0, \mathrm{y}_0, \mathrm{v}_{x_{0}}$ and $\mathrm{v}_{y_{0}}$. In these equations, the dynamic variables $(\mathrm{x}, \mathrm{y})$ represent the particle positions, while $(\mathrm{v}_x, \mathrm{v}_y)$ correspond to their velocities in the four-dimensional (4D) phase space. The governing system involves two independent non-dimensional parameters: the density parameter $\textrm{R} =2\rho_{p}/(2\rho_{p}+\rho_f)$ and the Stokes number $ \textrm{St} =\tau_{p}\,(U/L)$. As previously noted, the Stokes number $\textrm{St}$ characterizes particle inertia, where $\textrm{St} = 0$ corresponds to inertialess particles that perfectly trace the fluid streamlines. The density parameter $\textrm{R}$ quantifies the relative density of particles with respect to the fluid. Specifically, $\textrm{R} = 1$ corresponds to infinitely heavy particles ($\rho_p \gg \rho_f$), while $\textrm{R} = {2}/{3}$ represents neutrally buoyant particles ($\rho_p = \rho_f$). In contrast, $\textrm{R} = 0$ characterizes highly buoyant particles, such as bubbles ($\rho_p \ll \rho_f$). The coefficient of the history force, $\kappa$, is related to the other parameters as $\kappa = 3\,\sqrt{{\textrm{R}\,(1-\textrm{R})}/{\textrm{St}}}$. Note that our definition of the Stokes number, $\mathrm{St}$, differs from that used in other studies~\cite{daitche2013advection,prasath2019accurate,jaganathan2023basset}. These studies formulate $\mathrm{St}$ using the time scale $a^2/\nu$, representing the time scale over which momentum diffuses in the fluid. In contrast, our definition expresses the particle response time $\tau_p$, $(\rho_p/\rho_f)$ times than $a^2/\nu$, characterizing how quickly the particle's momentum adjusts to the surrounding fluid streamlines. For our analysis, we focus on the regime ${2}/{3} < \textrm{R} < 1$ and $0.1 < \textrm{St} < 1.0$, concentrating on the dynamics of particles that are finitely denser than the fluid and emphasize on investigating role of inertial and history forces on their transport features.

The governing system of equations (\ref{Eq:Equation_of_motion_TG_Flows}) consists of coupled nonlinear integro-differential equations, making analytical treatment highly challenging. Consequently, we resort to numerical methods to solve for particle dynamics. However, before proceeding with numerical simulations, we attempt to predict the particle behavior analytically by simplifying the flow field. The nonlinearity of the system primarily originates from the nonlinear nature of the TG vortex flow, which contains key flow structures such as vortex centers and stagnation points. We can study particle motion near these structures by linearising the flow in their vicinity to gain analytical insights. Of particular interest are the stagnation points, as previous studies~\cite{crisanti1990passive,wang1992chaotic} have shown that particles in a TG vortex flow primarily migrate between vortex cells through the regions surrounding these points. Specifically, heavy inertial particles ($\textrm{R}=1$) have been found to escape vortex cells near stagnation points if their Stokes number exceeds a critical threshold of $\textrm{St}=1/4$~\cite{nath2024irregular}. In the following section, we focus on how including inertial forces and the history force influences particle leakage near stagnation regions.
%%%%%%%%%%%%%%%%%%%%%%%%%%%%%%%%%%%%%%%%%%%%%%%%%%%%%%%%%%%%%%%%%%%%%%%%%%%%%%%%%%%%%%%%%%%%%%%
\section{\label{sec:ParticleDynamicsNearStagnantPoint}Particle Dynamics near a stagnation point}
The TG flow considered here has stagnation points at locations $(\mathrm{x},\mathrm{y}) = (n\,\pi, m\, \pi)$, where $n$ and $m$ are integers. For instance, when $n = m = 0$, the stagnation point is located at the origin. Each stagnation point in the TG vortex flow is surrounded by four vortices, as illustrated in Figure \ref{fig:Streamlines_TG_Flows}. The flow field near these stagnation points exhibits a pattern similar to that of a linear strain flow field (see the inset). For instance, near the stagnation point at the origin, the leading-order approximations for the TG vortex flow velocity components are $\textrm{u}_x \approx \textrm{x}$, and $\textrm{u}_y \approx -\textrm{y}$. While this velocity field specifically represents the linearised flow around the stagnation point at the origin, it can be generalized to describe the flow near any stagnation point in the TG vortex by appropriately accounting for the location and considering the extensional and compressional axes of the stagnation points. Before analyzing the full nonlinear system of the TG flow, we first examine particle dynamics in the stagnation flow region analytically. Initially, we only consider the effects of inertial forces, neglecting the history force. We then incorporate the history force to evaluate its impact on particle transport behavior.
%%%%%%%%%%%%%%%%%%%%%%%%%%%%%%%%%%%%%%%%%%%%%%%%%%%%%%%%%%%%%%%%%%%%%%%%%%%%%%%%%%%%%%%%%%%%%%%%%%%%%%%%%%%%%%%%%%%%%%%%%%%%%%%
\subsection{Without history force}
\label{sec3a}
\begin{figure*}[bth]
    \centering
\includegraphics[width=1\linewidth]{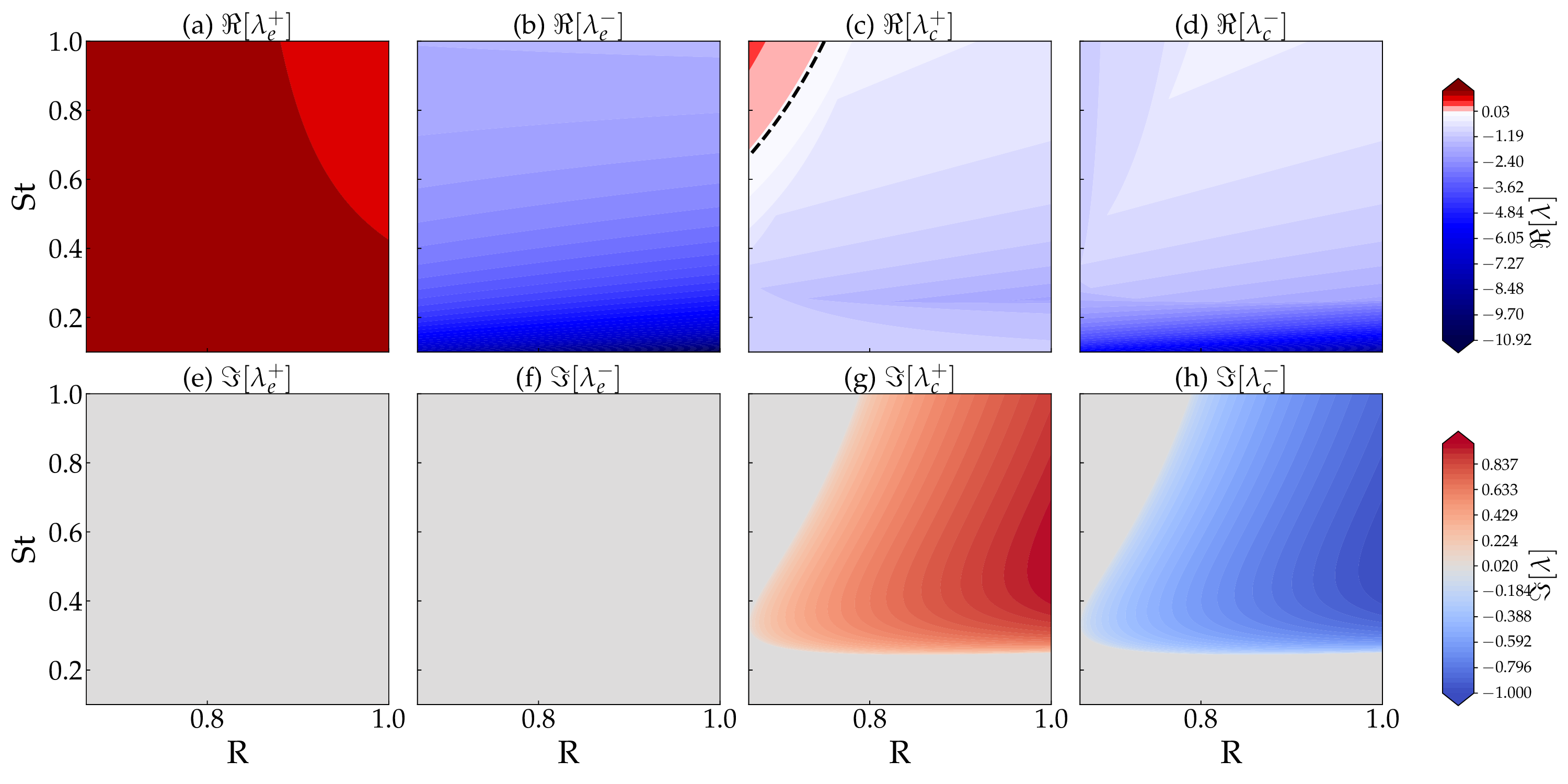}
    \caption{The real ($\Re$) and imaginary ($\Im$) parts of the eigenvalues $\lambda_{e}^{\pm}$ and $\lambda_{c}^{\pm}$ are presented as colored contour plots in the $\textrm{St}$--$\textrm{R}$ parametric plane. Zero contour lines are highlighted in black dashed in subplot (c)}
\label{fig:Real_and_imaginary_parts_eigenvalues}
\end{figure*}

Without loss of generality, we consider the stagnation point at the origin, where the governing equation (\ref{Eq:Equation_of_motion_TG_Flows}), neglecting the history force, simplifies to
\begin{subequations}
    \begin{align}
        \frac{1}{\textrm{R}}\ddot{\mathrm{x}} + \frac{1}{\textrm{St}}\dot{\mathrm{x}} - \mathrm{x}\left(\frac{3(1-\textrm{R})}{\textrm{R}}+\frac{1}{\textrm{St}}\right) = 0,\label{Eq:Without_History_Force_Linear_Flows_X} \\
         \frac{1}{\textrm{R}}\ddot{\mathrm{y}} + \frac{1}{\textrm{St}}\dot{\mathrm{y}} - \mathrm{y}\left(\frac{3(1-\textrm{R})}{\textrm{R}}- \frac{1}{\textrm{St}}\right) = 0.\label{Eq:Without_History_Force_Linear_Flows_Y}
    \end{align}    \label{Eq:Without_History_Force_Linear_Flows}
\end{subequations}
Here $\dot{(\,\, )}$ represents time derivative $d/dt$. Equations~(\ref{Eq:Without_History_Force_Linear_Flows_X}) and~(\ref{Eq:Without_History_Force_Linear_Flows_Y}) form a set of two decoupled dynamical systems, each representing an independent damped harmonic oscillator. The density parameter \textrm{R} effectively related to the mass of the oscillator, while the Stokes number \textrm{St} introduces damping effects. Both \textrm{R} and \textrm{St} also contribute to the spring constant of each oscillator. Note that the spring constants for the oscillators in $\mathrm{x}$ and $\mathrm{y}$ are slightly different. Since the equations of motion are decoupled, the solution for the compressional (y) and extensional (x) axes can be determined separately as %$\lambda_{c}^{\pm}$ and $\lambda_{e}^{\pm}$, respectively:}
\begin{widetext}
    \begin{subequations}
    \begin{align}
        \mathrm{x} = C_1\, e^{\lambda_{e}^+\, t}+C_2\, e^{\lambda_{e}^-\, t}, \quad \textrm{with}\,\, \,  \lambda_{e}^{\pm} = \frac{1}{2}\left(-\frac{\mathrm{R}}{\mathrm{St}} \pm \sqrt{\left(\frac{\mathrm{R}}{\mathrm{St}}\right)^2 + 4\frac{\mathrm{R}}{\mathrm{St}} + 12(1-\mathrm{R})}\right)~,       \label{Eq:Eigenvalues_extensional_axis}\\
            \mathrm{y} = C_3\, e^{\lambda_{c}^+\, t}+C_4\, e^{\lambda_{c}^-\, t}, \quad \textrm{with}\, \, \, \lambda_{c}^{\pm} = \frac{1}{2}\left(-\frac{\mathrm{R}}{\mathrm{St}} \pm \sqrt{\left(\frac{\mathrm{R}}{\mathrm{St}}\right)^2 - 4\frac{\mathrm{R}}{\mathrm{St}} + 12(1-\mathrm{R})}\right)~,
\label{Eq:Eigenvalues_compressional_axis}
\end{align}    \label{Eq:Eigenvalues_all}
\end{subequations}
\end{widetext}
 The constants $C_1, C_2, C_3$ and $C_4$ are integration constants and depend on the particle’s initial position and velocity as:
 \begin{widetext}
      \begin{equation}
         C_1 = \frac{\mathrm{v}_{x_{0}}-\mathrm{x}_0\, \lambda_{e}^-}{\lambda_{e}^+-\lambda_{e}^-}, \,C_2 = \frac{-\mathrm{v}_{x_{0}}+\mathrm{x}_0\, \lambda_{e}^+}{\lambda_{e}^+-\lambda_{e}^-}, \, C_3 = \frac{\mathrm{v}_{y_{0}}-\mathrm{y}_0\, \lambda_{c}^-}{\lambda_{c}^+-\lambda_{c}^-} , \textrm{and} \, \,  C_4 = \frac{-\mathrm{v}_{y_{0}}+\mathrm{y}_0\, \lambda_{c}^+}{\lambda_{c}^+-\lambda_{c}^-} ~.
         \label{C-Constants}
     \end{equation}
 \end{widetext}
 Here $\lambda_{e}^\pm$ and $\lambda_{c}^\pm$ are effectively the eigenvalues of the system of equations (\ref{Eq:Without_History_Force_Linear_Flows}) when treated as a linear dynamical system. The colormap in FIG.~\ref{fig:Real_and_imaginary_parts_eigenvalues} visualizes the real and imaginary parts of these eigenvalues across the $\textrm{St}$--$\textrm{R}$ parametric plane. The nature of particle trajectories can be inferred from the eigenvalue set ($\lambda_{e}^+,\lambda_{e}^-,\lambda_{c}^+,\lambda_{c}^-$), particularly from their signature.

 From the expression for $\lambda_{e}^\pm$, it is evident that these eigenvalues are always real since the discriminant $\mathfrak{D}_1=(\mathrm{R}/\mathrm{St})^2+4(\mathrm{R}/\mathrm{St})+12(1-\mathrm{R})$ remains positive for all physically relevant values of $ \mathrm{R} < 1 $. Consequently, $\lambda_{e}^+$ is always positive, while $\lambda_{e}^-$ is always negative, as illustrated in FIG.~\ref{fig:Real_and_imaginary_parts_eigenvalues}(a,b,e,f). The solution form in equation (\ref{Eq:Eigenvalues_extensional_axis}) thus indicates that particles will move exponentially away along the $\mathrm{x}$-axis (or, more generally, along the extensional axis). This behaviour can be understood by treating equation (\ref{Eq:Without_History_Force_Linear_Flows_X}) as a simple harmonic oscillator with a `negative spring constant'. While a negative spring constant is unphysical for a real oscillator, it theoretically implies a non-oscillatory, purely exponential response in the system's dynamics. Consequently, particle trajectories cannot intersect with the compressional axis in finite time, preventing them from moving across it—unless specific initial conditions enforce such a crossing, which will be discussed later in this section.

 In contrast, the eigenvalues $\lambda_{c}^\pm$ can be either real or complex, depending on parameter values, as shown in FIG.~\ref{fig:Real_and_imaginary_parts_eigenvalues}(c,d,g,h). When $\lambda_{c}^\pm$ are complex conjugates, the solution in equation (\ref{Eq:Eigenvalues_compressional_axis}) suggests that particle trajectories will exhibit oscillatory behavior along the $\mathrm{y}$-axis (or, in general, along the compressional axis). As a result, particles may able to cross the extensional axis in finite time, consistent with previous observations in \citet{nath2022transport}. To determine when $\lambda_{c}^\pm$ becomes complex, we consider the discriminant $\mathfrak{D}_2=(\mathrm{R}/\mathrm{St})^2-4(\mathrm{R}/\mathrm{St})+12(1-\mathrm{R})$. The eigenvalues become complex if the Stokes number lies within the range $\textrm{St}_{c}^+ < \textrm{St} < \textrm{St}_{c}^-$, where the critical Stokes numbers are given by $\textrm{St}_{c}^{\pm} = \textrm{R}/\{2(1\pm\sqrt{3\textrm{R}-2})\}$. For instance, in the heavy particle limit ($R = 1$), these reduce to $\textrm{St}_{c}^+ = {1}/{4}$ and $\textrm{St}_{c}^- \rightarrow \infty$, consistent with earlier studies \citep{taylor1940notes, Langmuir1944, nath2022transport}. When treating the system along the compressional axis as a simple harmonic oscillator, these critical Stokes numbers correspond to the condition of critical damping of the oscillator. For $\textrm{St}  \in (\textrm{St}_{c}^+,\textrm{St}_{c}^-)$, the oscillator is underdamped, resulting in sinusoidal variations in particle trajectories along the compressional axis. This oscillatory motion allows particles to cross the extensional axis without requiring additional conditions. The system transitions to an overdamped state for Stokes numbers beyond this critical range, where particle trajectories in phase space become purely exponential. In such cases, crossing can only occur under specific initial conditions, which will be discussed later in this section.
Additionally, whenever the eigenvalues $\lambda_{c}^\pm$ are real, $\lambda_{c}^{-}$ remains negative always, while $\lambda_{c}^{+}$ may assume positive values under specific parametric conditions. When both eigenvalues are negative, the trajectories converge toward the extensional axis at asymptotically large times. Conversely, if either eigenvalue is positive, such convergence cannot be anticipated. The expression for $\lambda_{c}^{+}$ indicates that it becomes positive if the Stokes number exceeds a critical threshold, defined by $\mathrm{St}_{c_{p}} = \mathrm{R}/\{3\,(1-\mathrm{R})\}$. This threshold effectively delineates two distinct regions in the parameter space, marking the transition of $\lambda_{c}^{+}$ from negative to positive values. In the context of the simple harmonic oscillator in equation (\ref{Eq:Without_History_Force_Linear_Flows_Y}), this regime corresponds to an oscillator with a `negative spring constant' along the compressional axis. Such a configuration implies particle trajectories diverge from the extensional axis (\textrm{y}=0 line). This diverging trajectory is already reported by~\citet{raju1997dynamics} calling it unphysical; Nevertheless, we propose that this particular parametric regime could facilitate particle transport across the extensional axis under appropriate initial conditions.

Now, let us consider the special case mentioned earlier, where particles can cross the extensional axis even if their trajectory is non-oscillatory, provided appropriate initial conditions are met. One might intuitively expect that for an inertial particle to cross a streamline, it simply needs sufficient momentum in the normal direction. The same principle applies here. Although the solution in equation (\ref{Eq:Eigenvalues_compressional_axis}) exhibits a non-oscillatory nature when both $\lambda_{c}^{+}$ and $\lambda_{c}^{-}$ are real, it is still possible for the initial conditions to result in $\mathrm{y} = 0$ at a finite time, indicating a chance for crossing the extensional axis. To determine the conditions for such a crossing, one can solve for $\mathrm{y}=0$ using equation (\ref{Eq:Eigenvalues_compressional_axis}), yielding the crossing time as $t_\textrm{cr} = \log(-C_4/C_3)/(\lambda_{c}^{+}-\lambda_{c}^{-})$. For this time to be a real, finite value, the conditions simplify to the following: when $\lambda_{c}^{+}$ and $\lambda_{c}^{-}$ are real, the particle trajectory can cross the compressional axis only if (i) for $\mathrm{y}_0 > 0$, the initial velocity satisfies $\mathrm{v}_{y_{0}}<\mathrm{y}_0\, \lambda_{c}^{-}$, and  (ii) for $\mathrm{y}_0 < 0$, the initial velocity satisfies $\mathrm{v}_{y_{0}}>\mathrm{y}_0\, \lambda_{c}^{-}$. Similarly, particle trajectories can also cross the compressional axis under appropriate initial conditions. Since the eigenvalues $\lambda_{e}^{+}$ and $\lambda_{e}^{-}$ are always real, the only way a particle can cross the compressional axis is through appropriate initial conditions. Using an identical approach but with equation (\ref{Eq:Eigenvalues_extensional_axis}), one can determine the conditions for $\mathrm{x} = 0$ to occur at a real, finite time in terms of the constants $C_1$ and $C_2$, leading to a similar criterion: the particle trajectory can cross the compressional axis only if (i) for $\mathrm{x}_0 > 0$, the initial velocity must satisfy $\mathrm{v}_{x_{0}}<\mathrm{x}_0\, \lambda_{e}^{-}$, and (ii) for $\mathrm{x}_0 < 0$, the initial velocity must satisfy $\mathrm{v}_{x_{0}}>\mathrm{x}_0\, \lambda_{e}^{-}$.

\begin{figure*}
    \centering
    \includegraphics[width=1.0\linewidth]{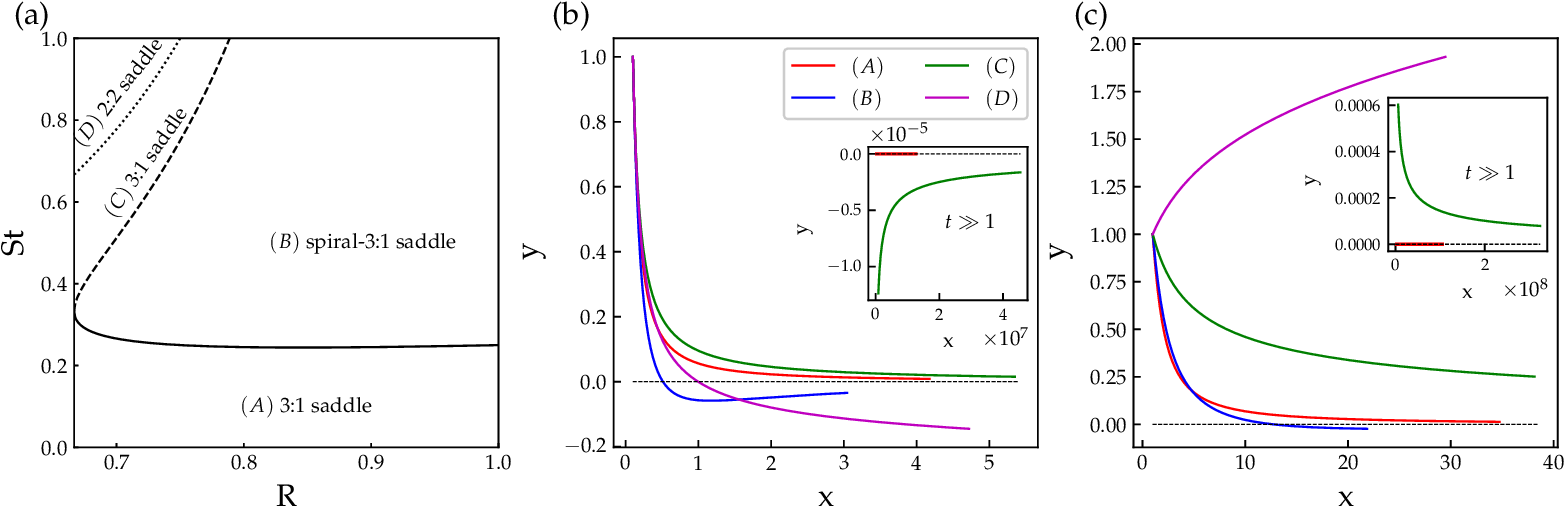}
    \caption{{(a) Different regimes of particle dynamics near a stagnation flow, without history force, are shown in the $\textrm{St}$--$\textrm{R}$ plane. The critical curves corresponding to $\textrm{St}_{c}^{+}$ (solid line), $\textrm{St}_{c}^{-}$ (dashed line), and $\mathrm{St}_{c_{p}}$ (dotted line) mark the boundaries between regions $A, B, C$ and $D$. Particle trajectories characteristic of each regime are illustrated in panel (b) for the case of zero initial slip velocity and panel (c) for the scenario of zero initial velocity. In both scenarios, the particles are initialized at the coordinate location (1,1), employing the following parameters: region $A$ with $\textrm{R}=0.8$, $\textrm{St}=0.2$; region $B$ with $\textrm{R}=0.85$, $\textrm{St}=0.5$; region $C$ with $\textrm{R}=0.67$, $\textrm{St}=0.45$; and region $D$ with $\textrm{R}=0.7$, $\textrm{St}=1.0$.}}   \label{fig:Critical_Stokes_plot_trajectories}
\end{figure*}

To summarise, based on the eigenvalue characteristics, we delineate distinct dynamical regions in the $\textrm{St}$--$\textrm{R}$ parameter space, as shown in FIG.~\ref{fig:Critical_Stokes_plot_trajectories}(a). Regions $A$ and $C$ exhibit similar behaviour, where all eigenvalues ($\lambda_{e}^+,\lambda_{e}^-,\lambda_{c}^+,\lambda_{c}^-$) are real with signatures $(+,-,-,-)$, corresponding to a `$3:1$ saddle' stagnation point in the 4D phase space. In region \( B \), bounded by \( \textrm{St}_{c}^{\pm} \), not all eigenvalues are purely real. The real parts retain the signature \( (+,-,-,-) \), while the imaginary parts have signatures \( (0,0,+,-) \), indicating a `spiral $3:1$ saddle’ in the phase space. In region $D$, beyond $\mathrm{St}_{c_{p}}$, all eigenvalues are real but have a different signature, $(+,-,+,-)$, corresponding to a `$2:2$ saddle’. The fixed point classification follows the nomenclature from \cite{hofmann2018visualization}. Notably, in region \( B \), unlike in \( A \), \( C \), and \( D \), the presence of complex-conjugate eigenvalues introduces a spiral nature to phase space trajectories. Specifically, these trajectories exhibit attracting spiral behaviour due to their negative real parts while retaining saddle-like properties owing to the remaining real eigenvalues. This spiral nature facilitates particle transport across the extensional axis in the stagnation flow region. In regions $A$, $C$, and $D$, particle trajectories can cross the extensional or compressional axis if they are initialised with sufficiently large momentum oriented appropriately. In the context of the TG vortex, these particle crossings may allow particles to traverse vortex cells in a finite time. 

To elucidate this concept, we present the trajectories of particles in physical space, specifically within the \( \textrm{x} \)–\( \textrm{y} \) plane, corresponding to various parametric regimes labelled \( A, B, C, \) and \( D \). We examine the implications of differing initial velocity conditions. Initially, we position the particles at the coordinates \( (1, 1) \) within a stagnation flow with zero initial slip and zero initial velocities in two distinct analyses. When the initial slip velocity is zero, the particle trajectories associated with regime \( A \) do not intersect the extensional axis (\( \mathrm{y}=0 \)) due to their non-oscillatory nature, ultimately aligning with the extensional axis over time, as illustrated in FIG.\ref{fig:Critical_Stokes_plot_trajectories}(b). In contrast, trajectories in regime \( B \) cross the extensional axis within a finite timeframe as a result of oscillatory motion along the compressional axis, as depicted in FIG.\ref{fig:Critical_Stokes_plot_trajectories}(b). Furthermore, in regime \( D \), the particle trajectory also intersects the extensional axis in finite time, as shown in FIG.~\ref{fig:Critical_Stokes_plot_trajectories}(b); however, this crossing is not attributable to oscillatory motion but rather due to specific initial conditions and momentum. Notably, particle trajectories associated with regime \( C \) cross the extensional axis at a later time \( (t \gg 1) \), as indicated in the inset of FIG.~\ref{fig:Critical_Stokes_plot_trajectories}(b). This behaviour occurs even without oscillatory features, underscoring the influence of initial velocity conditions. 

In a scenario where the initial velocity is zero, the trajectories in regimes \( A \) and \( B \) exhibit similar characteristics; specifically, the trajectories of the region \( A \) exhibit trapping behaviour, while those of region \( B \) cross the \( \mathrm{y}=0 \) line, as demonstrated in FIG.~\ref{fig:Critical_Stokes_plot_trajectories}(c). Trajectories associated with region \( C \) do not intersect the \( \mathrm{y}=0 \) line with zero initial velocity. This contrasts with previous observations where, with zero initial slip velocity, trajectories crossed the \( \mathrm{y}=0 \) line at late times \( (t \gg 1) \), as depicted in FIG.~\ref{fig:Critical_Stokes_plot_trajectories}(c). In the case of the region \( D \), particles directed against the flow do not cross the \( \mathrm{y}=0 \) line within a finite time, as shown in FIG.\ref{fig:Critical_Stokes_plot_trajectories}(c). This behaviour highlights the intricate relationship between initial conditions and particle trajectory dynamics in various regimes.

Recall that our analysis in this section focuses on particle dynamics in the vicinity of stagnation points in the TG vortex flow, where we approximate the flow as a simple stagnation flow through linearisation. Consequently, we should ideally consider small values of $\mathrm{x}$ and $\mathrm{y}$. However, as shown in insets of FIGs.~\ref{fig:Critical_Stokes_plot_trajectories}(b–c), the use of large values (e.g., on the order of $10^7$) extends beyond this intended context. These large values are primarily included to illustrate the behavior of particles in an unbounded straining flow. In the context of the TG vortex flow, by the time particles are predicted to cross the extensional or compressional axes, they may already be sufficiently far from the stagnation point such that the nonlinearity of the TG vortex flow can no longer be neglected. Consequently, the predictions may differ slightly when accounting for this nonlinearity. Furthermore, once a particle crosses a separatrix of the TG vortex flow, the nonlinear flow in the adjacent vortex cell may carry it further away from the stagnation point, with nonlinear effects ultimately governing its subsequent dynamics—potentially guiding it toward another stagnation region.
%%%%%%%%%%%%%%%%%%%%%%%%%%%%%%%%%%%%%%%%%%%%%%%%%%%%%%%%%%%%%%%%%%%%%%%%%%%%%%%%%%%%%%%%%%%%%%%%%
\subsection{With history force}
\label{sec3b}
Here, as an extension of the previous subsection, we consider particle motion in the stagnation flow while accounting for the history force. As before, the governing equations (\ref{Eq:Equation_of_motion_TG_Flows}) can be simplified in the vicinity of the stagnation point. Without loss of generality, we focus on the stagnation point at the origin of the TG vortex flow while retaining the history force, yielding the linear version of the governing equations as
\begin{widetext}
    \begin{subequations}
        \begin{align}
            \ddot{\mathrm{x}}+\frac{\textrm{R}}{\textrm{St}}\, \dot{\mathrm{x}}-\mathrm{x}\left({3\,(1-\textrm{R})}+\frac{\mathrm{R}}{\mathrm{St}}\right)=-\frac{\kappa}{\sqrt{\pi}}\left[\frac{\dot{\mathrm{x}}_0-{\mathrm{x}}_0}{\sqrt{t}}+\int_0^{t}\frac{1}{\sqrt{t-t'}}\left(\frac{d^2\mathrm{x}}{dt'^2}-\frac{d\mathrm{x}}{dt'}\right)dt'\right]~,\label{Eq:X_component_equation_with_history_linear_flow} \\
            \ddot{\mathrm{y}} +\frac{\textrm{R}}{\textrm{St}}\, \dot{\mathrm{y}}-\mathrm{y}\left({3\,(1-\textrm{R})}-\frac{\mathrm{R}}{\mathrm{St}}\right)=-\frac{\kappa}{\sqrt{\pi}}\left[\frac{\dot{\mathrm{y}}_0+{\mathrm{y}}_0}{\sqrt{t}}+\int_0^{t}\frac{1}{\sqrt{t-t'}}\left(\frac{d^2 \mathrm{y}}{dt'^2}+\frac{d\mathrm{y}}{dt'}\right)dt'\right]~.\label{Eq:Y_component_equation_with_history_linear_flow}
        \end{align}        \label{Eq:component_equations_with_history_linear_flow}
    \end{subequations}
\end{widetext}
These equations are integro-differential in nature and do not represent a simple dynamical system, like equations (\ref{Eq:Without_History_Force_Linear_Flows}). Although the left-hand sides of both systems remain identical, the right-hand sides differ due to forcing terms associated with the memory effect of the particle’s dynamical history. Here, this memory term appears as an integral. In the heavy particle limit ($\mathrm{R} \rightarrow 1$) or for large particle inertia ($\mathrm{St} \gg1$), both systems exhibit similar dynamical behavior since the coefficient of the history term, $\kappa$, approaches zero. Despite these complexities, the system in equations (\ref{Eq:component_equations_with_history_linear_flow}) remains linear, and the dynamics in the $\mathrm{x}$ and $\mathrm{y}$ directions remain decoupled, allowing analytical solutions through methods such as the classical Laplace transform. The solutions to equations~(\ref{Eq:component_equations_with_history_linear_flow}) in the Laplace domain are given by:
\begin{widetext}
\begin{subequations}\label{with_history_linear_flow_Laplace_domain}
        \begin{align}
            \hat{\mathrm{x}}=\frac{1}{\prod_{i=1}^{4}(\sqrt{s}-\mu_{x_{i}})}\left(s\;\mathrm{x}_0+{\kappa}\sqrt{s}\;\mathrm{x}_0+\frac{\textrm{R}}{\textrm{St}}\mathrm{x}_0+\dot{\mathrm{x}}_0\right),\label{X_with_history_linear_flow_Laplace_domain} \\
            \hat{\mathrm{y}}=\frac{1}{\prod_{i=1}^{4}(\sqrt{s}-\mu_{y_{i}})}\left(s\;\mathrm{y}_0+{\kappa }\sqrt{s}\;\mathrm{y}_0+\frac{\textrm{R}}{\textrm{St}}\mathrm{y}_0+\dot{\mathrm{y}}_0\right)~.\label{Y_with_history_linear_flow_Laplace_domain}
        \end{align}
        \label{Eqs10}
    \end{subequations}
\end{widetext}
Here, $\hat{\mathrm{x}}$ and $\hat{\mathrm{y}}$ are the Laplace transforms of the variables $\mathrm{x}$ and $\mathrm{y}$, respectively, while $\mu_{x_{i}}$ and $\mu_{y_{i}}$ are the roots  of the quartic polynomials $P_1(\xi)=\xi^4+{\kappa}\,\xi^3+({\textrm{R}}/{\textrm{St}})\,\xi^2-{\kappa}\,\xi-\left[{\textrm{R}}/{\textrm{St}}+{3(1-\textrm{R})}\right]$ and $P_2(\xi)=\xi^4+{\kappa}\,\xi^3+({\textrm{R}}/{\textrm{St}})\,\xi^2+{\kappa}\,\xi-\left[{3(1-\textrm{R})}-{\textrm{R}}/{\textrm{St}}\right]$. The roots $\mu_{x_{i}}$ and $\mu_{y_{i}}$ are independent of initial conditions and can be uniquely determined for given values of $\textrm{St}$ and $\textrm{R}$. Applying partial fraction decomposition and performing the inverse Laplace transform on equations~(\ref{Eqs10}), we obtain the general solutions for the particle trajectories $\textrm{x}$ and $\textrm{y}$ as
\begin{widetext}
    \begin{subequations}\label{trajectory_with_history_linear_flow}
        \begin{align}
            \mathrm{x}\left(t\right) = \Sigma_{i=1}^{4}A_{x_{i}}\mu_{xi}\exp{\left(\mu_{x_{i}}^2t\right)}\erfc\left(-\mu_{x_{i}}\sqrt{t}\right),\label{X_trajectory_with_history_linear_flow} \\
            \mathrm{y}\left(t\right) = \Sigma_{i=1}^{4}A_{y_{i}}\mu_{y_{i}}\exp{\left(\mu_{y_{i}}^2t\right)}\erfc\left(-\mu_{yi}\sqrt{t}\right)~.\label{Y_trajectory_with_history_linear_flow}
        \end{align}
    \end{subequations}
\end{widetext}
The coefficients $A_{x_{i}}$ and $A_{y_{i}}$ depend on the initial conditions, the roots $\mu_{x_{i}}$ and $\mu_{y_{i}}$, as well as the parameters \textrm{St} and \textrm{R}. They are given by
\begin{widetext}
    \begin{subequations}\label{coeff}
        \begin{align}
            A_{x_{i}}=\frac{\mathrm{x}_0\left(\mu_{x_{i}}^2+\kappa\mu{_{x_{i}}}\right)+\frac{\textrm{R}}{\textrm{St}}\mathrm{x}_0+\dot{\mathrm{x}}_0}{\prod_{{j\neq i},{j=1}}^{j=4}(\mu_{x_{i}}-\mu_{x_{j}})};\quad \Sigma_{i=1}^{4}A_{x_{i}}=0~, \\
            A_{y_{i}}=\frac{\mathrm{y}_0\left(\mu_{y_{i}}^2+\kappa\mu{_{y_{i}}}\right)+\frac{\textrm{R}}{\textrm{St}}\mathrm{y}_0+\dot{\mathrm{y}}_0}{\prod_{{j\neq i},{j=1}}^{j=4}(\mu_{y_{i}}-\mu_{y_{j}})};\quad \Sigma_{i=1}^{4}A_{y_{i}}=0~.
        \end{align}
    \end{subequations}
\end{widetext}
The complementary error function $(\erfc)$ in the general solution, along with the exponential terms, captures the influence of the history force. Analyzing the roots of the characteristic polynomials $P_1(\xi)$ and $P_2(\xi)$ provides insights into particle trajectories, though this can be complex. For large times ($t \gg 1$), particle trajectories generally tend towards the extensional axis (here, $\mathrm{y} = 0$) at a rate of approximately $\sim t^{-1/2}$, unless the real parts of the roots $\mu_{y_{i}}$ are significantly large and positive. This implies that, compared to cases without history effects, particles approach a line parallel to $\mathrm{y} = 0$ more gradually. However, whether the trajectories cross $\mathrm{y} = 0$ remains uncertain and requires further analysis. Determining this would involve solving equation~\ref{Y_trajectory_with_history_linear_flow} for the time at which $\mathrm{y(t)} = 0$, but this is analytically challenging and is not pursued here. Instead, a numerical investigation of this aspect is presented later in this section.

\begin{figure}
    \centering
    \includegraphics[width=1.0\linewidth]{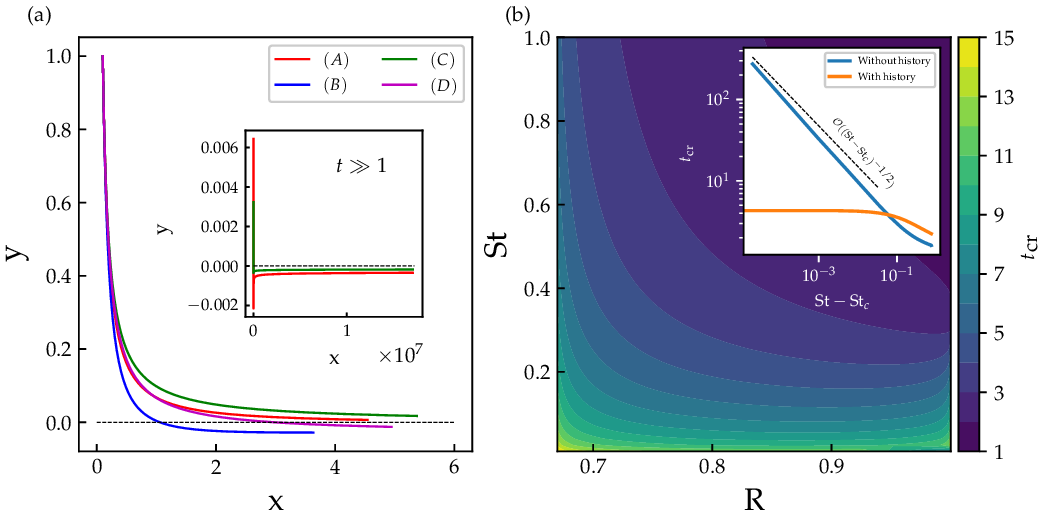}
    \caption{(a) Typical particle trajectories under the influence of the history force. Particles are initialised at location ($1,1$) with zero initial slip velocity for different parameter combinations: region $A$ with $\textrm{R}=0.8$, $\textrm{St}=0.2$; region $B$ with $\textrm{R}=0.85$, $\textrm{St}=0.5$; region $C$ with $\textrm{R}=0.67$, $\textrm{St}=0.45$; and region $D$ with $\textrm{R}=0.7$, $\textrm{St}=1.0$. The inset shows particle trajectories at $t\gg 1$ for regions $A$ and $C$. (b) Coloured contour plot of the escape time $T_\textrm{esc}$, representing the time taken for a particle initialised at $(1,\;1)$ with zero slip velocity to cross the extensional axis ($\mathrm{y}=0$), is shown in the \textrm{St}--\textrm{R} parameter space, accounting for the history force. The inset plot of the panel (b) compares the variations in critical crossing time, denoted as \( t_{\mathrm{cr}} \), with \( \mathrm{St} - \mathrm{St}_{c} \), under scenarios both with and without the influence of history force. The resultant curve obtained in the absence of history force exhibits a power law scaling behaviour characterized by the relationship \( (\mathrm{St} - \mathrm{St}_{c})^{-1/2} \).}
    \label{fig:Escape_time_history_force}
\end{figure}

To examine the effect of the history force on particle trajectories, we present representative trajectories for different parameter combinations, all initialised at location $(1,1)$ near the stagnation zone with zero initial slip velocity, as shown in FIG.~\ref{fig:Escape_time_history_force}(a). The parameter combinations used here are the same as in FIG.~3(b) and (c), but only the zero initial slip velocity case is considered. The selected trajectories correspond to the parametric conditions from distinct regions $(A)$, $(B)$, $(C)$, and $(D)$ of the $\textrm{St}$--$\textrm{R}$ parameter space, as discussed in the previous section. Similar to the observations in the absence of the history force, particles originating from regions $(B)$, $(C)$, and $(D)$ cross the extensional axis ($\mathrm{y}=0$ line) within finite time, as seen in FIG.~\ref{fig:Escape_time_history_force}(a). In region $(D)$, particles with and without the history force cross the separatrix. However, unlike in the absence of history effects, those influenced by the history force do 
show exponential divergence from the extensional axis. Here, the history force counteracts this divergent behaviour introduced by the added mass and pressure gradient forces. Additionally, an important difference here is observed in region $(A)$: with the history force, particles cross the $\mathrm{y}=0$ line, a phenomenon not seen when the history force is absent.

To determine whether particles can cross the extensional axis under the influence of the history force, we adopt a numerical approach due to the complexity of an analytical solution. Using the solution in equation (\ref{Y_trajectory_with_history_linear_flow}), we initialize particles at $\mathrm{y}_0 = 1$ with zero slip velocity and track whether they cross the $\mathrm{y}=0$ line numerically. If a particle crosses the axis, we record the first occurrence of such an event and define it as the critical escape time, $t_\textrm{cr}$. This process is repeated for a range of parameter combinations, and the results are presented as a colored contour plot in the \textrm{St}--\textrm{R} plane in FIG.~\ref{fig:Escape_time_history_force}(b). The results show that for all parameter combinations considered, particles eventually cross the extensional axis in a finite time. However, the escape time varies with \textrm{St} and \textrm{R}, with shorter escape times in some regions (indicated by blue) and longer times in others (yellow). Specifically, particles with low \textrm{St} values and \textrm{R} close to $0.67$ take significantly longer to escape compared to those with high \textrm{St} values and \textrm{R} near $1$, as evident from FIG.~\ref{fig:Escape_time_history_force}(b). Additionally, as $\textrm{R}\to 1$, the escape time is large for $\textrm{St}<1/4$ but reduces for $\textrm{St}>1/4$, consistent with results in the absence of the history force. Notably, unlike the case without history effects, no subregions within the parameter space indicate permanent trapping of particles near the stagnation point when the history force is considered. As the Stokes number ($\mathrm{St}$) approaches the critical value $\mathrm{St}_{c}$ at a given ($\mathrm{R}$), the critical escape time diverges in the absence of history force. Conversely, the curve representing the critical escape time ($t_{\mathrm{cr}}$) manifests a smooth variation in the presence of the history force. The inset in panel FIG.\ref{fig:Escape_time_history_force}(b) illustrates a comparison of the $t_{\mathrm{cr}}$ curve as it varies with $\mathrm{St} - \mathrm{St}_{c}$, both with and without the consideration of history force. It is observed that the divergence of $t_{\mathrm{cr}}$ in the absence of history force near the critical Stokes number $\mathrm{St}_{c}$ adheres to a $-1/2$ scaling law with respect to $\mathrm{St} - \mathrm{St}_{c}$, assuming that the particles are initialized with zero slip velocity. Away from $\mathrm{St}_{c}$, in the context of heavy particles, the variation of $t_{\mathrm{cr}}$ when accounting for the history force follows a similar trend with respect to $\mathrm{St}$ as that observed without the history force, albeit with a larger magnitude, as demonstrated in the inset of FIG.~\ref{fig:Escape_time_history_force}(b).
%%%%%%%%%%%%%%%%%%%%%%%%%%%%%%%%%%%%%%%%%%%%%%%%%%%%%%%%%%%%%%%%%%%%%%%%%%%%%%%%%%%%%%%%%%%%%%%%%
\section{\label{sec:Particle dynamics in TG vortex flow}Particle dynamics in the TG vortex flow - without history force}
In this section, we analyze the dynamics of inertial particles in TG vortex flows while neglecting the history force but accounting for other forces, such as the pressure gradient force and added mass effects. The equations of motion for an inertial particle in the TG vortex flow, in the absence of the history force, can be derived from equations (5) by omitting the terms multiplying to $\kappa$, as:
\begin{widetext}
\begin{subequations}
\begin{align}
\label{X_Equation_without_history_TG_flow}
 \dot{\mathrm{x}} = \mathrm{v}_{\mathrm{x}},\quad \dot{\mathrm{v}}_{\mathrm{x}} = - \frac{\textrm{R}}{\textrm{St}}\,\left(\mathrm{v}_x-\sin{\mathrm{x}}\,\cos{\mathrm{y}}\right) + {3\,(1-\textrm{R})}\,\sin{\mathrm{x}}\,\cos{\mathrm{x}}~,\\
 \label{Y_Equation_without_history_TG_flow}
     \dot{\mathrm{y}} = \mathrm{v}_\mathrm{y},\quad \dot{\mathrm{v}}_\mathrm{y} = - \frac{\textrm{R}}{\textrm{St}}\,\left(\mathrm{v}_y + \sin{\mathrm{y}}\, \cos{\mathrm{x}}\right) + {3\,(1-\textrm{R})}\,\sin{\mathrm{y}}\,\cos{\mathrm{y}}~.
     \end{align}
     \label{eq13}
\end{subequations}
\end{widetext}
These equations are coupled and nonlinear but remain ordinary differential equations. They can be rewritten by defining a modified effective flow field that incorporates the modifications introduced by the fluid inertia term, allowing it to be absorbed into the Stokes drag as:
\begin{widetext}
\begin{subequations}
\begin{align}
\label{X_equation_with_added_mass_effective_flow}
    \dot{\mathrm{x}} &= \mathrm{v}_x, \quad \dot{\mathrm{v}}_x = -\frac{\textrm{R}}{\textrm{St}}\,\left(\mathrm{v}_x - \mathrm{u}_{xe}\right)~,\\
\label{Y_equation_with_added_mass_effective_flow}
    \dot{\mathrm{y}} &= \mathrm{v}_y, \quad \dot{\mathrm{v}}_y = -\frac{\textrm{R}}{\textrm{St}}\,\left(\mathrm{v}_y - \mathrm{u}_{ye}\right)~.
     \end{align}
     \label{eq14}
\end{subequations}
\end{widetext} 
The modified flow field is given by $ \mathrm{u}_{xe} = \sin{\mathrm{x}}\, (\cos{\mathrm{y}}+\gamma^{-1}\cos{\mathrm{x}}) $ and $ \mathrm{u}_{ye} = \sin{\mathrm{y}}\,(- \cos{\mathrm{x}}+\gamma^{-1}\cos{\mathrm{y}}) $, where $\gamma=\mathrm{St}_{c_{p}}/\textrm{St}$ and, as defined earlier, $\mathrm{St}_{c_{p}} = \mathrm{R}/\{3\,(1-\mathrm{R})\}$. The modified flow field differs significantly from the conventional TG vortex flow due to its dependence on $\textrm{St}$ and $\textrm{R}$, meaning that different particles, depending on their size and density, can experience different effective flow fields. Interestingly, the modified flow field $\mathbf{u}_{e}=[\mathrm{u}_{xe}, \mathrm{u}_{ye}]^{T}$ is compressible, with a nonzero divergence given by $\nabla \cdot \mathbf{u}_e = \gamma^{-1}(\cos{2\mathrm{x}} + \cos{2\mathrm{y}})$. In the heavy particle limit ($\textrm{R}\to 1$), the modified flow field converges to the conventional TG vortex flow, and the modified equations of motion (equations~(\ref{X_equation_with_added_mass_effective_flow}) and~(\ref{Y_equation_with_added_mass_effective_flow})) reduce to the simplified Maxey–Riley equation~(\ref{Simplified_Maxey-Riley_equation}) for TG vortex flow.

\begin{figure}
    \centering   \includegraphics[width=1.0\linewidth]{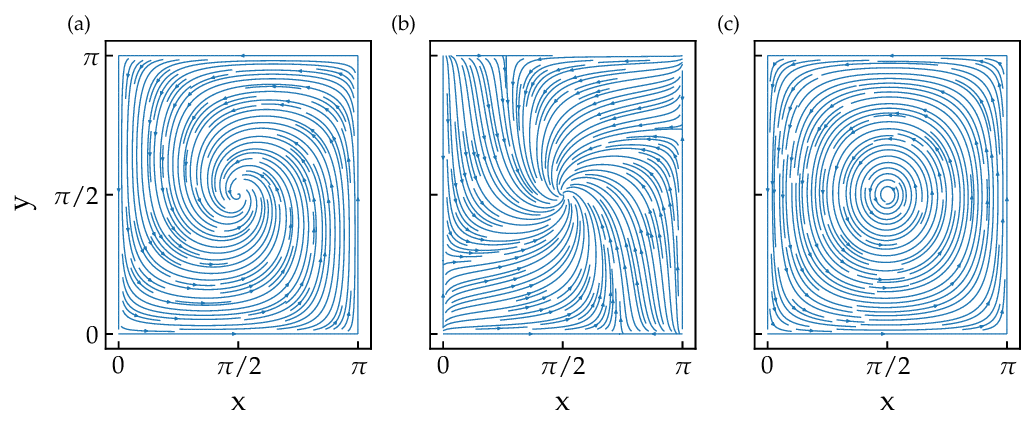}
    \caption{Streamlines of the effective flow field $[\mathrm{u}_{xe}, \mathrm{u}_{ye}]^{T}$ are shown for different values of  $ \textrm{R} $ and $ \textrm{St} $. (a) For $ \textrm{R} = 0.67 $ and $ \textrm{St} = 0.3 $, the streamlines exhibit a spiralling sink behaviour at the vortex centre, with no additional fixed points. (b) When $\textrm{R}=0.67$ and $\textrm{St}=1.0$, four new saddle-type fixed points present. (c) For $\textrm{R}=0.99$ and $\textrm{St}=1.0$, the flow structure closely resembles the classical TG vortex pattern.}
    \label{fig:StreamPlot}
\end{figure}

By setting $ \dot{\mathrm{x}} = 0 $, $ \dot{\mathrm{y}} = 0 $, $ \dot{\mathrm{v}}_x = 0 $, and $ \dot{\mathrm{v}}_y = 0 $, we can determine the fixed points of particles in this effective flow field. Here also trivially the solution gives $\mathrm{v}_x = \mathrm{v}_y = 0$. In addition to the regular TG flow fixed points at ($n\, \pi, m\, \pi$), ($n\, \pi+\pi/2, m\, \pi+\pi/2$), we find additional fixed points located at ($n\, \pi, \cos^{-1}((-1)^n\, \gamma)$), ($\cos^{-1}(-(-1)^m\, \gamma), m\, \pi$), for any integers $n$ and $m$. For instance, in the basic vortex cell bounded by the separatrices at $\mathrm{x}=0, \mathrm{x}=\pi, \mathrm{y}=0$ and $\mathrm{y}=\pi$, four additional fixed points appear apart from the vortex centre and the four corner stagnation points. These new fixed points are located at ($0,\cos^{-1}(\gamma)$), ($\pi,\cos^{-1}(-\gamma)$), ($\cos^{-1}(\gamma), \pi$), and ($\cos^{-1}(-\gamma),0$). These are saddle points, as can be visualized in, for example, FIG.~\ref{fig:StreamPlot}(b). Notably, these additional fixed points fall on the flow separatrices, but their exact location depends on $\gamma$, meaning their locations vary based on particle size and density. The existence of these fixed points is restricted to cases where $\gamma < 1$, which corresponds to $\textrm{St}>\mathrm{St}_{c_{p}}$. In the heavy particle limit ($\textrm{R}\to 1$), since $\mathrm{St}_{c_{p}} \rightarrow \infty$, no particle with a finite Stokes number would perceive these new fixed points. However, for any finite density ratio ($\textrm{R}< 1$), there exists critical Stokes number $\mathrm{St}_{c_{p}}$ beyond which these additional fixed points emerge. Interestingly, this critical Stokes number coincides with $\mathrm{St}_{c_{p}}$, despite being introduced earlier in an entirely different context in the case of stagnation flow. At present, the underlying reason for this coincidence remains unclear. FIG.\ref{fig:StreamPlot} illustrates the streamlines of the effective flow field $\mathbf{u}_{e}$ in the basic vortex cell, highlighting the emergence of additional fixed points for different combinations of $ \textrm{R} $ and $ \textrm{St} $. When  $ \textrm{R} = 0.67$ and $ \textrm{St}  = 0.3 < \mathrm{St}_{c_{p}} $, the streamlines exhibit a spiralling pattern characteristic of sink flow at the vortex centre, with no additional fixed points, as shown in FIG.\ref{fig:StreamPlot}(a). However, when the Stokes number increases to $ \textrm{St} =1 > \mathrm{St}_{c_{p}} $, four additional fixed points emerge, as depicted in FIG.\ref{fig:StreamPlot}(b). As mentioned earlier, they can be qualitatively identified as saddles types. In contrast, for a nearly heavy particle of $ \textrm{R} = 0.99 $ with $\textrm{St} =1 < \mathrm{St}_{c_{p}}$, the spiralling behaviour is diminished, and the streamlines closely resemble the conventional TG vortex pattern, as shown in FIG.\ref{fig:StreamPlot}(c).

A formal linearization and eigenvalue analysis of the effective flow field $[\mathrm{u}_{xe}, \mathrm{u}_{ye}]^{T}$ near the fixed points can provide insights into the local behaviour of streamlines. Near the corner fixed points, the eigenvalues of the effective flow field are given by $(1+\gamma^{-1})$ and $(-1+\gamma^{-1})$, indicating a saddle-type behaviour when $\textrm{St}<\mathrm{St}_{c_{p}}$, while the behaviour transition to source-type when $\textrm{St}>\mathrm{St}_{c_{p}}$. The source-type behaviour of corner fixed points are possible here because of the compressible nature of the effective flow field. In the scenario when $\textrm{St}>\mathrm{St}_{c_{p}}$, the additional fixed points emerge with eigenvalues $(\gamma+\gamma^{-1})$ and $(\gamma-\gamma^{-1})$, exhibiting a saddle-type behaviour. Additionally, the fixed point at the cell center has eigenvalues $(-\gamma^{-1}\pm i)$, signifying a spiral attractor when $\gamma^{-1} \neq 0$. It is important to note that these eigenvalues and the associated fixed point classifications describe the nature of streamlines in the effective flow field near their vicinity, which aligns well with the examples in FIG.\ref{fig:StreamPlot}. However, the effective flow field is merely a construct that represents the modified flow perceived by an inertial particle. Despite this representation, inertial particles will behave differently than streamlines in this effective flow field due to their finite inertia. To completely understand inertial particle dynamics, one must linearize the full governing system of equations, such as Eqs.~(\ref{eq13}) or Eqs.~(\ref{eq14}), about the fixed points in the phase space. Accounting for this, the stability of the corner fixed points follows the same pattern as discussed in the context of stagnation flow in \S \ref{sec3a}. Similarly, the stability of inertial particles near the vortex centre can be determined, showing that they behave as a `2-spiral saddle' when $\textrm{R} > 2/3$ and as a `2-spiral sink' when $\textrm{R}<2/3$. Since here we consider denser particles ($\textrm{R} > 2/3$), they spiral outward from the vortex centre despite the stable spiral behaviour observed in the effective flow streamlines at the cell centre. This means that although the effective flow field suggests an inward spiral at the cell centre, inertial particle trajectories actually exhibit outward spiralling motion. Similar to the case of the old fixed points, the dynamics of an inertial particle near the new fixed points are analyzed in their vicinity by linearizing the flow and using linear stability analysis in the following subsection.
%%%%%% 
%%%%%%%%%%%%%%%%%%%%%%%%%%%%%%%%%%%%%%%%%%%%%%%%%%%%%%%%%%%%%%%%%%%%%%%%%%%%%%%%%%%%%%%%%%%%
\subsection{\label{sec:additional stagnant points}Particle dynamics near to a new fixed point of modified flow field}
\label{sec4a}
As mentioned earlier, once the new fixed points exist, they exhibit a saddle-type nature from the perspective of inertial particles. Consequently, they possess extensional and compressional axes, which would be appropriately switched for other new fixed points. Therefore, without loss of generality, as in \S \ref{sec3a}, we can focus on one of the new fixed points, let us say the one at $ \left(\cos^{-1}\left(-\gamma\right), 0\right) $ in the basic vortex cell. We linearize equations (\ref{X_equation_with_added_mass_effective_flow}) and (\ref{Y_equation_with_added_mass_effective_flow}) about this fixed point, yielding the following simplified governing equations:
\begin{equation}\label{X_equation_with_added_mass_effective_flow_linearised}
     \frac{\textrm{St}}{\textrm{R}}\,\ddot{\mathrm{x}}' +\dot{\mathrm{x}}' -  \left(\gamma-\gamma^{-1}\right)\, \mathrm{x}'=0~,
\end{equation}
\begin{equation}\label{Y_equation_with_added_mass_effective_flow_linearised}
     \frac{\textrm{St}}{\textrm{R}}\,\ddot{\mathrm{y}}' +\dot{\mathrm{y}}' -  \left(\gamma+\gamma^{-1}\right)\, \mathrm{y}'=0~, 
\end{equation}
where $\mathrm{x}'$ and $\mathrm{y}'$ represent the coordinates measured relative to the fixed point location. As expected for an inertial particle in any saddle-like flow field, the system reduces to two independent linear harmonic oscillators. Consequently, the solution can be expressed as a linear combination of exponential functions, similar to Eqs.~(\ref{Eq:Eigenvalues_all}), albeit with different eigenvalues and integration constants. The eigenvalues in this case are:
\begin{widetext}
\begin{subequations}
\begin{align}
\label{eq17a}
    \lambda_{1}^{\pm}=\frac{1}{2\, \textrm{St}}\, \left[-\textrm{R} \pm \sqrt{\textrm{R}\, \left(\textrm{R}+4\, \textrm{St}\, (\gamma-\gamma^{-1})\right)} \right]~,\\
\label{eq17b}
     \lambda_{2}^{\pm}=\frac{1}{2\, \textrm{St}}\, \left[-\textrm{R} \pm \sqrt{\textrm{R}\, \left(\textrm{R}+4\, \textrm{St}\, (\gamma+\gamma^{-1})\right)} \right]~.
     \end{align}
     \label{eq17}
\end{subequations}
\end{widetext}
Similar to the case of corner fixed points in \S \ref{sec3a}, here, the eigenvalues corresponding to the new fixed points also exhibit a transition in behaviour within the parametric plane. The eigenvalues $\lambda_{2}^{\pm}$ remain purely real for all relevant parameter regimes, with $\lambda_{2}^{+}$ always positive and $\lambda_{2}^{-}$ always negative. However, the eigenvalues $\lambda_{1}^{\pm}$ can be purely real and negative for certain parameter regimes but transition into complex conjugate pairs with a negative real part when the Stokes number exceeds a critical value. This critical Stokes number can be determined by setting the discriminant of $\lambda_{1}^{\pm}$ in Eq.~(\ref{eq17a}) to zero, yielding ${\textrm{R}+4\, \textrm{St}\, (\gamma-\gamma^{-1})} = 0$. Solving for \textrm{St}, we obtain the critical Stokes number as $\textrm{St}_{c_2} = (\textrm{R}\, \sqrt{7 - 3\,\textrm{R}})/(6\, (1-\textrm{R}))$. Consequently, the nature of inertial particle trajectories near the new fixed points in phase space corresponds to a 3:1 saddle when $\textrm{St}<\textrm{St}_{c_2}$, transitioning to a spiral-3:1 saddle when $\textrm{St}>\textrm{St}_{c_2}$. It is also important to note that these new fixed points exist only if $\textrm{St}>\mathrm{St}_{c_{p}}$. Similar to the case of corner fixed points, the emergence of spiral characteristics in the phase-space trajectories suggests enhanced particle leakage in the vicinity of these new fixed points when $\textrm{St}>\textrm{St}_{c_2}$.

\begin{figure}
    \centering
    \includegraphics[width=1.0\linewidth]{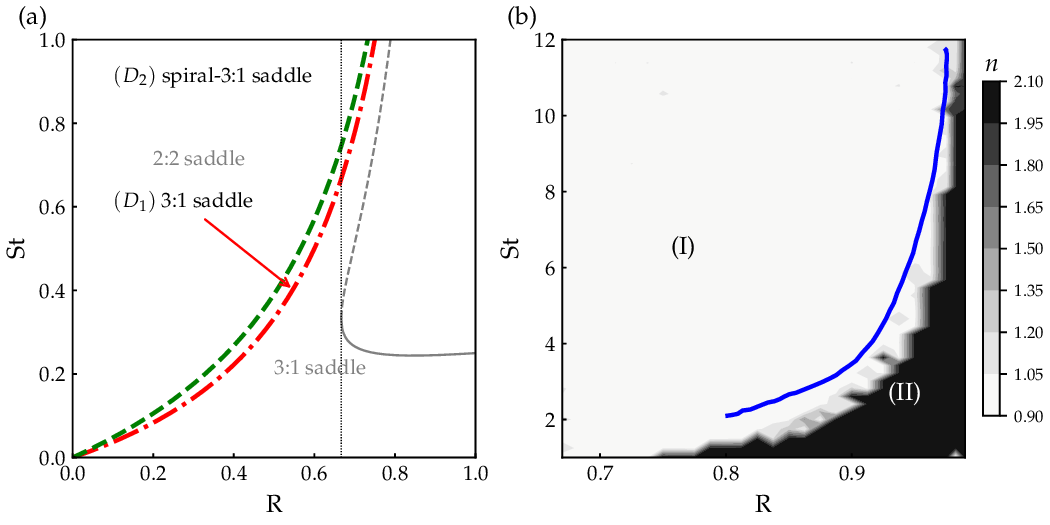}
    \caption{(a)~The figure illustrates the critical curve corresponding to $\mathrm{St}_{c_{p}}$ in red (dashed-dotted), marking the threshold above which region $D$ emerges, where new fixed points appear. The green (dashed) curve, corresponding to $\textrm{St}_{c_2}$, further divides this region into sub-regions $D_1$ and $D_2$. For comparison, the other critical curves that separate regions $A, B$ and $C$ from $D$, associated with the corner fixed points, are shown in the background in gray. (b)~The regions for chaotic and periodic particle motion in the parameter space of $\textrm{St}$ and $\textrm{R}$ from the study by~\citet{wang1992chaotic}. Region (I) corresponds to chaotic particle motion, while Region (II) corresponds to periodic particle motion. The red curve represents the boundary identified in their study, which is corrected in our study, as evident from the contour plot.}
    \label{fig:Modified_Stokes_Critical_Wang_Maxey}
\end{figure}

In FIG.~\ref{fig:Modified_Stokes_Critical_Wang_Maxey}(a), the red dashed-dotted curve represents the critical line $\mathrm{St}_{c_{p}} = \mathrm{R}/\{3\,(1-\mathrm{R})\}$ in the $\textrm{St}$--$\textrm{R}$ parametric plane, marking the threshold above which new fixed points emerge. Interestingly, this region coincides with region $D$ in FIG.~\ref{fig:Critical_Stokes_plot_trajectories}(a). This region is further subdivided into two sub-regions, $D_1$ and $D_2$, by the green dashed curve corresponding to $\textrm{St}_{c_2} = (\textrm{R}\, \sqrt{7 - 3\,\textrm{R}})/(6\, (1-\textrm{R}))$. As mentioned earlier, in region $D_1$, the inertial particle trajectories near the new fixed points in the 4D phase space exhibit a 3:1 saddle behaviour, whereas in region $D_2$, they transition to a spiral-3:1 saddle. Although our primary interest lies in the range $\textrm{R} \in (2/3,1)$, the parametric plot has been extended to the broader range $\textrm{R} \in (0,1)$ to illustrate the generality of our results. The regions \(A\) and \(B\) merge into a single domain for lighter particles \((\mathrm{R}<2/3)\). In this framework, the old corner fixed points display a 3:1 saddle behaviour as long as the Stokes number stays below the critical threshold \(\mathrm{St}_{c_p}\). When the Stokes number exceeds \(\mathrm{St}_{c_p}\), the dynamics of these fixed points shift to a 2:2 saddle configuration. FIG.~\ref{fig:Modified_Stokes_Critical_Wang_Maxey}(a) graphically illustrates this behaviour, where grey colour marks the regions, and fixed point dynamics correspond to the old fixed points. For lighter particles, the new fixed points remain active for Stokes numbers greater than \(\mathrm{St}_{c_p}\), showcasing 3:1 saddle characteristics. As the Stokes number increases beyond \(\mathrm{St}_{c_2}\), these characteristics transform into a spiral 3:1 saddle configuration, as seen in FIG.~\ref{fig:Modified_Stokes_Critical_Wang_Maxey}(a).
%%%%%%%%%%%%%%%%%%%%%%%%%%%%%%%%%%%%%%%%%%%%%%%%%%%%%%%%%%%%%%%%%%%%%%%%%%%%%%%%%%%%%%%%%%%%%%%%%%%
\section{Particle dynamics in the full nonlinear TG vortex flow – without history force}
\label{sec5}
In \S \ref{sec3a} and \S \ref{sec:additional stagnant points}, we analysed the dynamics of inertial particles near the fixed points of the TG vortex flow and examined the critical criteria associated with particle leakage across flow separatrices in the absence of the history force. The governing dynamical equations became linear since the flow field was locally linear near these fixed points, making the analysis analytically tractable and relatively straightforward. However, once particles leak across the flow separatrices, they traverse the nonlinear TG vortex flow structures before encountering another fixed point. Hence, a comprehensive study of the full nonlinear system is necessary to understand the complete particle dynamics. Due to the nonlinear nature of the governing equations, an analytical approach becomes challenging, necessitating numerical methods. In this section, we investigate the dynamics of inertial particles in the TG vortex flow without history force effects, while considering the pressure gradient force and added mass effects, by numerically solving the governing equations (\ref{eq13}). As these equations form a set of coupled, nonlinear ordinary differential equations, we employ a fourth-order Runge–Kutta scheme for numerical integration.

To utilize the periodic nature of the flow field, we initialize $10^4$ particles uniformly within the basic vortex cell $[0, \pi]\times[0, \pi]$. We initialize them with zero initial slip velocities and track their trajectories by integrating the governing equations~(\ref{X_Equation_without_history_TG_flow}) and~(\ref{Y_Equation_without_history_TG_flow}) from $t=0$ to $t=10^{4}$ nondimensional time units. This study specifically focuses on the dynamics of denser particles rather than lighter ones. Consequently, the density parameter $\textrm{R}$ is varied within the range $2/3$ to $1$, while the Stokes number is restricted to values less than or equal to $1$ for simplicity. To characterize the mean behaviour of the particles, we compute the mean (averaged over different initializations) square displacement (MSD), defined as
\begin{equation}
    \textrm{MSD} = \langle \; \lVert\mathbf{x}(t)-\mathbf{x}_{0}\rVert^{2} \; \rangle~,
\end{equation}
Here, $\langle \, \cdot \,\rangle$ denotes the average over all particle trajectories $\mathbf{x}(t)$ originating from respective initial positions $\mathbf{x}_0$, while \( \lVert \, \cdot \, \rVert \) represents the $2$-norm. The long-time scaling behaviour of $\textrm{MSD}$ with time, given by $\textrm{MSD} \sim t^{n}$, characterizes the collective particle dynamics. If particle dispersion is dominated by diffusion, the exponent is $n=1$; if ballistic motion dominates, $n=2$; and if particles exhibit limited dispersion due to trapping, $n=0$. A previous study by \citet{wang1992chaotic} examined the dynamics of inertial particles in the TG vortex flow, excluding history effects, and identified regions of regular and chaotic motion in the $\textrm{St}$--$\textrm{R}$ parameter space. In their analysis, they employed an older form of the added mass term, similar to that in original Maxey-Riley equation~\cite{maxey1983equation}, which effectively introduces an additional term, $(1-\textrm{R})\,(\mathbf{v}-\mathbf{u})\cdot \boldsymbol{\nabla} \mathbf{u}$ to the equations of motion in Eq.~(\ref{Maxey_Riley equation}), thereby altering the form of Eqs.~(\ref{eq13}). The authors considered particles with $\textrm{St}\lesssim 12$ and $\textrm{R}$ in the range $(0.8,0.975)$, reporting that particles with higher $\textrm{St}$ and lower $\textrm{R}$ exhibit diffusive dispersion at large times due to chaotic (random walk like) trajectories. Conversely, particles in other regimes display periodic motion, leading to either open or closed trajectories, which correspond to ballistic or trapped dispersion, respectively. To assess the influence of the correct form of the added mass term on particle dynamics and dispersion, we first reproduce the results of \citet{wang1992chaotic} using our corrected governing equations (\ref{eq13}) for the TG vortex flow. This allows us to examine the impact of the proper added mass formulation on long-time particle dispersion. For direct comparison, we also transform their results to align with the parameter definitions of $\textrm{R}$ and $\textrm{St}$ used in this study.

\citet{wang1992chaotic} numerically obtained the boundary separating regular and chaotic regimes in the parametric plot. We have digitized this boundary from their figure 3 and included it as a blue continuous curve here in FIG.\ref{fig:Modified_Stokes_Critical_Wang_Maxey}(b). Additionally, FIG.\ref{fig:Modified_Stokes_Critical_Wang_Maxey}(b) presents a colour-coded representation distinguishing regions of regular and chaotic particle dynamics based on the corrected added mass term from our study. To achieve this, we conducted simulations over a $26 \times 54$ discretized grid of $\textrm{R} \times \textrm{St}$ values to compute their $\textrm{MSD}$ and the corresponding exponent value $n$ from it. The resulting contour plot of $n$ in FIG.~\ref{fig:Modified_Stokes_Critical_Wang_Maxey}(b) reveals that the boundary separating regular and chaotic regions now differs significantly from the blue curve reported by~\citet{wang1992chaotic}. Furthermore, this new boundary appears highly irregular. The figure also suggests that the earlier study by \citet{wang1992chaotic} underestimated the extent of the chaotic region for lower $\textrm{R}$ values $(\sim 2/3)$, as we observe chaotic motion even for lower $\textrm{St}$ values (e.g., $\textrm{St} < 2$) in this regime. In the limit $\textrm{R}\to 1$, the results obtained using the old and corrected forms of the added mass term are almost identical, as the additional term $(1-\textrm{R})\,(\mathbf{v}-\mathbf{u})\cdot \boldsymbol{\nabla} \mathbf{u}$ from the old formulation becomes negligible; even the added mass effect itself diminishes.

\begin{figure}
    \centering
\includegraphics[width=1.0\linewidth]{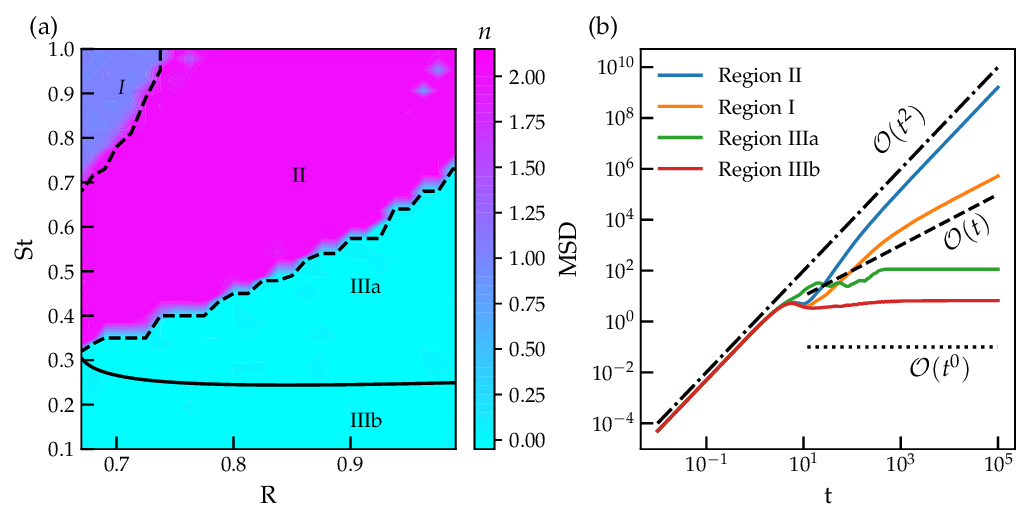}
    \caption{(a) Regions of diffusive (I), ballistic (II) and trapped (IIIa and IIIb) motion on the parametric plot of $\textrm{St}-\textrm{R}$. (b) MSD versus time plot corresponding regions I\;(\textrm{R}=0.67,\;0.9), II\;(\textrm{R}=0.7,\;0.95, IIIa\;(\textrm{R}=0.99,\;0.7) and IIIb\;(\textrm{R}=0.7,\;0.1)}
    \label{fig:Variance_contour}
\end{figure}

In FIG.~\ref{fig:Variance_contour}(a), we present a contour plot of $n$ in the $\textrm{St}$--$\textrm{R}$ parameter space, similar to FIG.\ref{fig:Modified_Stokes_Critical_Wang_Maxey}(b), but with a focus on the  $\textrm{St}<1$ range, where more interesting particle dynamics are observed. We identify three distinct regimes in the parameter space, each corresponding to different long-time particle behaviours based on the scaling of \textrm{MSD} with time. The region \textrm{I} in this parametric plot is an extension of the region \textrm{I} in FIG.\ref{fig:Modified_Stokes_Critical_Wang_Maxey}(b), representing a diffusive regime where particles exhibit chaotic, random-walk-like dispersion. This regime is less pronounced in FIG.\ref{fig:Variance_contour}(a), as it is mostly confined to \(\textrm{R} \lesssim 0.75\) and \(\textrm{St} \gtrsim 0.7\), but it becomes more dominant over a broader $\textrm{St}$ range in FIG.~\ref{fig:Modified_Stokes_Critical_Wang_Maxey}(b). Similarly, region \textrm{II} in FIG.\ref{fig:Variance_contour}(a) extends region \textrm{II} from FIG.\ref{fig:Modified_Stokes_Critical_Wang_Maxey}(b), where particle motion is dominantly ballistic. However, unlike FIG.\ref{fig:Modified_Stokes_Critical_Wang_Maxey}(b), FIG.\ref{fig:Variance_contour}(a) reveals an additional region, denoted as region \textrm{III}, where particle dispersion is limited, resulting in $n=0$. This region signifies particles that either remain trapped by flow separatrices and stagnation regions or are confined in periodic limit cycle trajectories. The curve corresponds to $\textrm{St}_{c}^{+}$  divides region \textrm{III} into two subregions. The lower part, the region \textrm{IIIb}, is the same as region $A$ in FIG.~\ref{fig:Critical_Stokes_plot_trajectories}(a), where particles remain trapped within their initial vortex cell, constrained by separatrices and stagnation regions. Conversely, in region \textrm{IIIa}, particles may escape to neighbouring vortex cells before becoming trapped by their separatrices or stagnation regions. Alternatively, their trajectories may evolve into periodic limit cycles, confining them to specific bounded, closed paths. In both cases, MSD eventually saturates to a constant value at large times. FIG.~\ref{fig:Variance_contour}(b) presents the MSD for representative $\textrm{St}$ and $\textrm{R}$ values corresponding to these different regimes. The MSD for region \textrm{I} scales as $t$, characteristic of diffusive motion, while in region \textrm{II}, it scales as $t^2$, indicating ballistic transport. In contrast, MSD in regions \textrm{IIIa} and \textrm{IIIb} saturates, though the saturation value is higher in region \textrm{IIIa}, suggesting that particles travel farther from the initial vortex cell before becoming trapped.

Our findings differ from those of~\citet{wang1992chaotic} in two key ways. They reported that in TG vortex flow, diffusion of particles due to their chaotic motion happens only for $\textrm{St} > 2$ with $\textrm{R} \in (0.8,1)$, whereas we find diffusion starting at $\textrm{St} > 0.7$ even when $\textrm{R} \approx 0.67$. Also, while they observed a relatively smooth boundary between regular and chaotic regimes in the parametric space, we find these boundaries to be highly irregular. These differences may be due to their use of an incorrect added mass term, which could have affected the results. Another reason could be the limited numerical support and the smaller parameter space they explored. 
Our study also focuses only on denser particles and explores the parameter range $\textrm{R} \in (2/3,1)$. It is possible that similar interesting particle dynamics occur in the regime $\textrm{R} \in (0,2/3)$, which we have not investigated in this study. We also find that the boundary between region \textrm{II} (ballistic) and region \textrm{III} (trapped) approaches $0.77$ as $\textrm{R} \rightarrow 1$, consistent with recent findings by~\citet{nath2024irregular}. Beyond this threshold, particle trajectories transition from a bounded state to an unbounded one for heavy inertial particles.

When analyzing individual particle trajectories, particularly in region \textrm{II}, we found that some particles exhibit diffusive motion at large times, even though the MSD identifies this regime as ballistic. This pattern also appears in other regions, where MSD predicts a dominant behaviour, but not all particles follow the same large-time dynamics. Instead, some particles exhibit some other behaviour, such as diffusion or trapping, depending on their initial positions and velocities. For instance, even if MSD suggests that particle dispersion should be ballistic for a given $\textrm{St}$ and $\textrm{R}$ (e.g., in region $\textrm{II}$), not all particles with these parameter values will necessarily exhibit ballistic motion. When ballistic particles are present, the MSD averages over them, suppressing the contributions of particles following subdominant dynamics (diffusive or trapped), making them less apparent. This highlights that MSD only captures the dominant mean behaviour, whereas individual particles may follow different trajectories. Such behaviour indicates the non-ergodic nature of the system. A similar observation was also reported by \citet{nath2024irregular} for heavy inertial particles in a TG vortex flow. Our study confirms their findings and extends them to finitely dense particles. In an ergodic system, a single particle’s dynamics would match the system’s mean behaviour, allowing mean dynamical studies to be sufficient. However, the absence of ergodicity here means that individual particle dynamics cannot be fully described by MSD alone. Instead, single-particle squared displacement (SD) must be examined to determine how it scales with time at large times.

\begin{figure*}
    \centering
\includegraphics[width=1.0\linewidth]{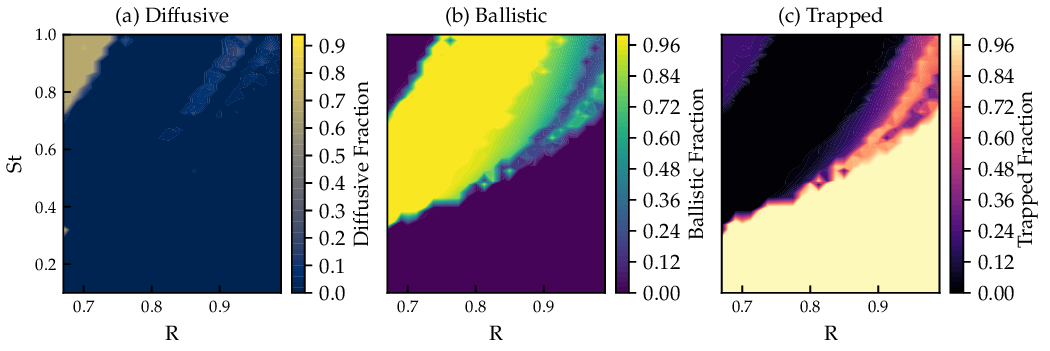}
    \caption{Fraction of particles exhibiting different types of dynamics: (a) diffusive, (b) ballistic, and (c) trapped, shown as coloured contour plots in the $\textrm{St}-\textrm{R}$ parameter space. A total of $10^4$ particles are uniformly initialized within the basic vortex cell with zero initial slip velocity, for every $\textrm{St}$--$\textrm{R}$ combination. Their single-particle SD is evaluated, and its scaling with time at the large time limit is used to classify their dynamics.}
\label{fig:Contour_Fraction_Particle_Without_History}
\end{figure*}

Following the classification approach of \citet{nath2024irregular}, we categorize particles, uniformly initialized within the basic vortex cell with zero initial slip velocity, into trapped, diffusive, and ballistic groups based on their long-term dispersion behaviour. The fractions of each category are shown as contour plots in the $\textrm{St}$--$\textrm{R}$ parameter space in FIG.~\ref{fig:Contour_Fraction_Particle_Without_History}(a–c). FIG.~\ref{fig:Contour_Fraction_Particle_Without_History}(a) indicates that diffusive motion is primarily confined to region $\textrm{I}$, with some isolated patches elsewhere. FIG.~\ref{fig:Contour_Fraction_Particle_Without_History}(b) shows that ballistic motion dominates region $\textrm{II}$, though its lower boundary is highly irregular. FIG.~\ref{fig:Contour_Fraction_Particle_Without_History}(c) reveals that most trapped particles are in the region $\textrm{III}$, though some also appear in regions $\textrm{I}$ and $\textrm{II}$. Notably, a narrow strip of trapped particles exists in region \textrm{II}, particularly near its lower boundary.

%%%%%%%%%%%%%%%%%%%%%%%%%%%%%%%%%%%%%%%%%%%%%%%%%%%%%%%%%%%%%%%%%%%%%%%%%%%%%%%%%%%%%%%%%%%%
\section{\label{sec:NumeicalResults}The effect of the history force on particle dynamics in the fully nonlinear TG vortex flow}
%Particle dynamics in the full nonlinear TG vortex flows}
In the previous sections, \S \ref{sec3a} and \S \ref{sec4a}, we examined the dynamics of inertial particles near fixed points in the TG vortex flow through linearization, while the full TG vortex flow without history effects was analyzed in \S \ref{sec5}. The influence of the history force on particles only near saddle points in the TG vortex flow (or a general stagnation flow) was considered in \S \ref{sec3b}. In this section, we conduct a detailed investigation of the effect of history force on inertial particle dynamics in the fully nonlinear TG vortex flow. Specifically, we explore how the particle dynamics observed in \S \ref{sec5} are modified when the history force term is included. To achieve this, we solve the full system of equations (\ref{Eq:Equation_of_motion_TG_Flows}). Since an analytical approach as in \S \ref{sec3b} is not feasible because of the non-linearity, we employ numerical methods for this analysis.

\begin{figure*}
    \centering
\includegraphics[width=1.0\linewidth]{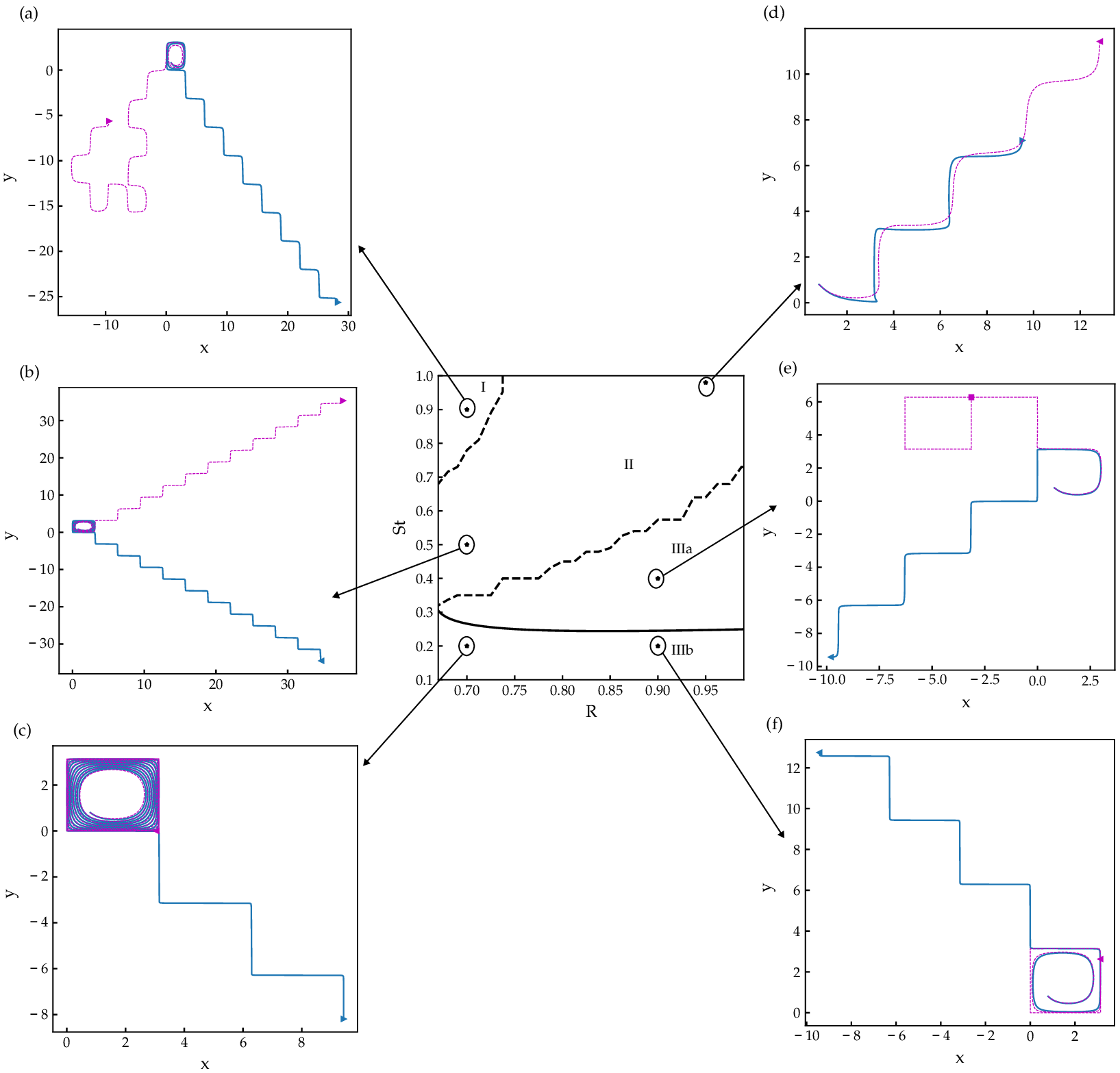}
    \caption{Typical particle trajectories in the presence of the history force (blue, continuous) are compared with those in its absence (magenta, dashed) for different combinations of $\textrm{St}$ and $\textrm{R}$. The particles are initialized with identical positions $(0.8,\;0.8)$ and velocities (zero slip velocity) in the TG vortex flow. The central subfigure marks all parametric combinations considered in the $\textrm{St}$--$\textrm{R}$ plane, which also highlights dynamical regime boundaries (in the absence of the history force) identified in earlier sections with black curves. Subfigures (a–f) compare particle trajectories with and without the history force for the following parameter values: (a) $\textrm{R} = 0.7$, $\textrm{St} = 0.9$, (b) $\textrm{R} = 0.7$, $\textrm{St} = 0.5$, (c) $\textrm{R} = 0.7$, $\textrm{St} = 0.2$, (d) $\textrm{R} = 0.95$, $\textrm{St} = 0.98$, (e) $\textrm{R} = 0.9$, $\textrm{St} = 0.4$, and (f) $\textrm{R} = 0.9$, $\textrm{St} = 0.2$. Arrows at the trajectory endpoints indicate that the particle remains untrapped and continues moving, while square markers denote particles that have been trapped at the respective locations.}
    \label{fig:Typical particle trajectories with history force}
\end{figure*}

Including the history force introduces several challenges, even in the numerical integration of the full Maxey–Riley equation. These challenges stem from the singularity of the kernel $ {1}/{\sqrt{t - t'}} $, at the upper limit of integration ($ t' = t $), and the substantial memory demands due to the history dependence of the integration process. Over the past few decades, researchers have developed various numerical algorithms to address these issues when integrating the Maxey–Riley equation with the history force~\cite{michaelides2022multiphase,alexander2004high,van2011efficient,daitche2013advection,elghannay2016development,prasath2019accurate,jaganathan2023basset}. In this section, we numerically integrate our governing equations (\ref{Eq:Equation_of_motion_TG_Flows}) using the algorithm proposed in Ref.~\cite{jaganathan2023basset}. In the previous section \S~\ref{sec5}, we identified regions of different particle dynamics in the $\textrm{St}$--$\textrm{R}$ parameter space in the absence of the history force, as illustrated in Fig.~\ref{fig:Variance_contour}(a). Here, we analyze the influence of the history force on particle dynamics in these regions. Fig.~\ref{fig:Typical particle trajectories with history force} presents a comparative study, where the central subfigure outlines the different regimes identified in Fig.~\ref{fig:Variance_contour}(a), while subfigures (a–f) depict single-particle trajectories for representative values of $\textrm{St}$ and $\textrm{R}$ from these regions. Each case includes trajectories both with and without the history force for direct comparison. In all cases, the particles are initialized at identical positions $(0.8,\; 0.8)$ with the same initial velocities, which is zero slip velocity. The left subfigures (a–c) correspond to particles that are slightly denser than the fluid, whereas the right subfigures (d–f) represent cases where the particles are significantly denser.

For $\textrm{R} = 0.7$ and $\textrm{St} = 0.2$, as predicted in \S \ref{sec3a}, trapped dynamics is observed when the history force was absent. However, upon including the history force, the particle is no longer confined within the vortex cell, as shown in FIG.\ref{fig:Typical particle trajectories with history force}(c). This behaviour is consistent with our findings in \S \ref{sec3b}, where the inclusion of the history force was shown to disrupt the critical criterion for particle leakage. For $\textrm{R} = 0.7$ and $\textrm{St} = 0.5$, both with and without the history force, the particle escapes the vortex cell and exhibits ballistic motion at large times, as seen in FIG.\ref{fig:Typical particle trajectories with history force}(b). However, despite identical initial conditions, the two trajectories diverge: the trajectory without the history force follows an average direction of $45^\circ$, whereas the trajectory with the history force is directed towards $-45^\circ$, with the horizontal. This deviation arises due to the history force, which we hypothesize gets amplified when the particle traverses stagnation regions. For $\textrm{R} = 0.7$ and $\textrm{St} = 0.9$, particles without the history force escape the vortex cell and follow chaotic, random-walk-like trajectories, whereas the inclusion of the history force results in regular ballistic motion, as shown in FIG.\ref{fig:Typical particle trajectories with history force}(a).

Next, we consider the case where particles are significantly denser than the fluid ($\textrm{R} \geq 0.9$). When $\textrm{R} = 0.9$ and $\textrm{St} = 0.2$, a particle without the history force remains trapped in the vortex cell. However, in the presence of the history force, it escapes, as seen in FIG.\ref{fig:Typical particle trajectories with history force}(f). This behaviour is similar to that in FIG.~\ref{fig:Typical particle trajectories with history force}(c), even though the influence of the history force is minimal here in the limit of heavy particles ($\textrm{R} \approx 1$). Also, the spiral part of the trajectory is wider here due to larger particle inertia. For $\textrm{R} = 0.9$ and $\textrm{St} = 0.4$, a particle without the history force exits its initial vortex cell but becomes trapped in another vortex cell. In contrast, when the history force is included, the particle continues to move without confinement across many vortex cells, as illustrated in FIG.\ref{fig:Typical particle trajectories with history force}(e). Finally, for $\textrm{R} = 0.95$ and $\textrm{St} = 0.98$, both cases (with and without history) exhibit regular ballistic motion, with trajectories traversing multiple vortex cells, as shown in FIG.~\ref{fig:Typical particle trajectories with history force}(d). Ballistic trajectories have identical angles from the horizontal, while trajectories using history force approach the fixed points more closely than those without it.

\begin{figure}
    \centering \includegraphics[width=0.5\linewidth]{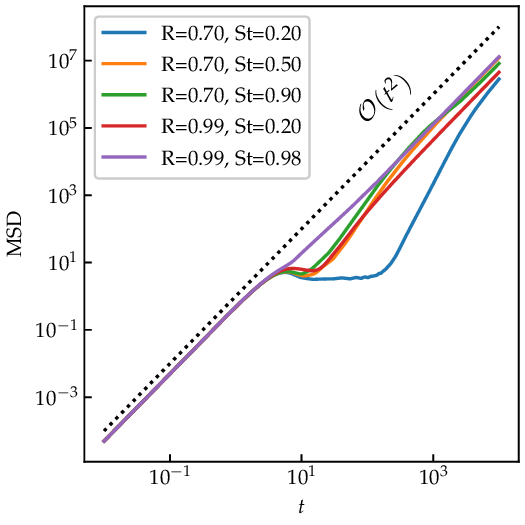}
    \caption{MSD versus time plot for a few representative cases of $\textrm{St}$ and $\textrm{R}$. All particles are initially distributed uniformly within the basic vortex cell with zero initial slip velocity.}
    \label{fig:MSD and exponent with history force}
\end{figure}

We report that the inclusion of the history force significantly alters particle dynamics. In this case, particle trapping by flow separatrices no longer occurs, as predicted in \S \ref{sec3b}. Additionally, all particles exhibit ballistic motion in the large-time limit, irrespective of $\textrm{St}$, $\textrm{R}$, or their initial locations. This behavior has been verified for two cases of initial particle velocity: (i) zero initial velocity, and (ii) zero initial slip velocity. For further clarification, FIG.~\ref{fig:MSD and exponent with history force} presents the MSD computed over $10^4$ particle trajectories, initially distributed within the basic vortex cell with zero initial slip velocity, for several representative cases of $\textrm{St}$ and $\textrm{R}$ as a function of time. All MSD curves asymptotically follow a $ t^{2} $ scaling at large times, confirming the presence of long-term ballistic motion. For verification, we also generated a contour plot of the exponent $n$, as in FIG.~\ref{fig:Variance_contour}(a), but we incorporated the effects of the historical force. However, the plot was not included, as it displayed a uniform colour, indicating that the numerically computed $n$ values were close to 2 for all cases, with slight deviations. We attribute these deviations to the finite simulation time, which was restricted to $t = 10^4$. If the simulations were extended to much larger times, we expect that $n$ would asymptotically approach $2$.
From FIG.~\ref{fig:MSD and exponent with history force}, it is also evident that while the MSD asymptotically approaches a ballistic regime, some cases exhibit transient trapped behaviour at intermediate times. This shows a residual influence of the dynamics observed in the absence of the history force.
This study is dedicated to highlighting the effects of the history force on particle dynamics and how they differ from cases where only inertial forces (pressure gradient and added mass effects) are considered along with Stokes drag. However, the earlier sections focus on particle dynamics in the absence of the history force, as these cases are also relevant in certain physical scenarios and cannot simply be dismissed as oversimplified approximations. In our analysis, we have employed a specific form of the history force corresponding to the Boussinesq-Basset history kernel, which decays as $\mathcal{O}(t^{-1/2})$ with time. Our findings indicate that this force significantly influences particle dynamics. In particular, it allows particles to cross flow separatrices even at low  $\textrm{St}$, $\textrm{R}$, in finite time. Furthermore, at asymptotically large times, the presence of the history force results in ballistic transport for all particles, irrespective of values of $\textrm{St}$, $\textrm{R}$ or initial conditions. However, these modifications due to the history force must be interpreted with caution. The Boussinesq-Basset kernel is valid in the limit of $Re = 0$. For any finite but small Reynolds number ($Re \ll 1$), more appropriate history kernels exist~\cite{lovalenti1993hydrodynamic,mei1994flow,dorgan2007efficient}, which exhibit a much faster decay, such as $\mathcal{O}(t^{-2})$. If such a kernel were used, the long-time ballistic behaviour observed in this section would not be an accurate representation of the true dynamics. Instead, at sufficiently large times, the influence of the history force would diminish relative to other inertial forces, making the dynamical behaviours described in the previous sections (\S \ref{sec3a}, \S \ref{sec4a}, and \S \ref{sec5}, where the history force was neglected purposely) more appropriate. Nevertheless, for short times, the history force remains relevant and should not be excluded from consideration.
%%%%%%%%%%%%%%%%%%%%%%%%%%%%%%%%%%%%%%%%%%%%%%%%\
\section{\label{sec:Conclusion}Discussion and conclusion}
We investigated the dynamics of inertial particles that are denser than fluid in a 2D steady TG vortex flow, characterized by an infinite array of counter-rotating vortices. Tracers remain confined within the vortical cells, following the closed streamlines of the flow. Conversely, dense inertial particles with $\textrm{St} \neq 0$ and $\textrm{R} > 2/3$, exhibit the potential to cross streamlines and traverse through these vortical structures, depending on the specific values of \textrm{St} and \text{R}. Previous studies have reported that under the heavy particle limit, where $\textrm{R}=1$, inertial particles remain trapped within a vortex cell if the Stokes number satisfies the condition $\textrm{St} < \mathrm{St}_{c} = 1/4$, assuming that the impacts of inertial and history forces are negligible.

We derived a general expression for the critical Stokes number \(\mathrm{St}_{c}\) by employing linear stability analysis to examine particle dynamics in the vicinity of stagnation points of TG vortex flow. We categorized the distinct regions of dynamical behaviour on the \(\textrm{St}\)--\(\textrm{R}\) parameter plane based on the characteristics of the eigenvalues.  In region \(A\), situated below the critical Stokes number line \(\mathrm{St}_{c}^{+}\), particles exhibited trapping behaviour as the stagnation point showed 3:1 saddle features. As a result, the particle trajectories are restricted to move across the flow separatrices, leading to particle trapping. Notably, this trapping dynamics remained consistent irrespective of whether the particles were initialized with zero velocity or zero slip. In region \(B\), bounded by the curves \(\mathrm{St}_{c}^{+}\) and \(\mathrm{St}_{c}^{-}\), particles traversed the stagnation point, characterized by a transition of the stagnation point to a spiral-3:1 saddle configuration. Here, the oscillatory nature of particle trajectories allows particles to move across the flow separatrices and lead them to neighboring vortex cells. Region \(C\), delineated by \(\mathrm{St}_{c}^{-}\) and \(\mathrm{St}_{c_{p}}\), shared similarities in stagnation point features with region \(A\). However, particle dynamics in this region showed a strong dependence on the initial conditions of the particles. Even though the particle trajectories do not have an oscillatory nature, appropriate initial velocities can make them cross the flow separatrices. Finally, the region \(D\), located beyond \(\mathrm{St}_{c_{p}}\), was characterized by particle trajectories that deviated from the extensional axis. Here the fixed point has a 2:2 saddle nature in the phase space. In this region also, the ability of particles to cross the stagnation point depended on their initial velocities. We presented an analysis regarding the emergence of additional stagnation points in the flow field, as perceived by the inertial particles in an effective flow field, attributing these to the influence of inertial forces (added mass and pressure gradient forces). We linearized the flow field in proximity to these new fixed points also and analyzed the corresponding dynamics of the particles, as well as the trajectory equations. Notably, these newly identified stagnation points only exist in the region \(D\) of the parametric space. We also examined the eigenvalue characteristics to delineate region \(D\) into two subregions: \(D_1\), where these new fixed points exhibited 3:1 saddle behaviour alongside already existing corner fixed points which are 2:2 saddles; and \(D_2\), where the behaviour of the new fixed points transitioned to a spiral-3:1 saddle nature, while the old corner fixed points remained as 2:2 saddles.

In the absence of a history force, the critical time \( t_{\mathrm{cr}} \) required for the particles to cross the extensional axis of the flow field \( (\mathrm{y}=0) \) exhibits a singular scaling \( (\mathrm{St} - \mathrm{St}_c)^{-1/2} \) as the Stokes number \( \mathrm{St} \) approached the critical Stokes number \( \mathrm{St}_c \) for a given density parameter \( \mathrm{R} \). Conversely, in the presence of a history force, the \( t_{\mathrm{cr}} \) curve demonstrated a smooth variation with \( \mathrm{St} \), indicating that particles are able to cross the extensional axis, irrespective of the values of \( \mathrm{St} \) and \( \mathrm{R} \), provided they are initialised with zero slip velocity. For zero initial velocity of the particles also we expect qualitatively the same behaviour to exit. A coloured contour map of \( t_{\mathrm{cr}} \) in the parametric space of \( \mathrm{St} \) and \( \mathrm{R} \) illustrated that particles characterized by lower \( \mathrm{St} \) and \( \mathrm{R} \) values required a longer duration to cross the extensional axis, whereas particles with higher \( \mathrm{St} \) and \( \mathrm{R} \) values required comparatively less time for crossing.

We numerically investigated particle dynamics in the fully nonlinear TG vortex flow, focusing particularly on large-time dispersion behaviour, which we quantified using MSD. We identified and delineated parameter regimes in the $\textrm{St}$-$\textrm{R}$ plane, where trapped, diffusive, and ballistic particle dynamics emerged at the large time limit. Our analysis revealed two distinct trapped regions in the parameter space: the first corresponded to particles remaining confined within their initial vortex cell, similar to the trapping zone observed in the linearized flow cases, while the second corresponded to particles becoming trapped in a different vortex cell after crossing the initial one.

For a given parametric combination of $\textrm{St}$ and $\textrm{R}$, we tracked individual particle trajectories and identified their dynamical nature based on the scaling of their single-particle squared displacement (SD) with time. This scaling gave us insight into individual particle dispersion behaviour, going beyond the mean trend indicated by the MSD. Our analysis revealed that while the MSD reflected the dominant behaviour of particles initialized within the vortex cell, not all particles exhibited the same dynamics. Even when the MSD indicated ballistic behaviour, a fraction of particles displayed sub-dominant diffusive or trapped dynamics. Accordingly, we quantified and reported the fractions of particles demonstrating ballistic, diffusive, and trapped behaviour. This observation aligned with a previous study~\cite{nath2024irregular} in the heavy particle limit.

In the presence of the history force, our investigation revealed modifications to particle dynamics in the nonlinear TG vortex flow. We concentrated our analysis on the restricted parameter regime: $\textrm{St}\in[0.1,1]$ and $\textrm{R} \in (2/3,1)$. The introduction of the history force led to the disappearance of the previously identified trapped and diffusive regions in the $\textrm{St}$--$\textrm{R}$ parametric space. Notably, particles characterized by low values of $\textrm{St}$ and $\textrm{R}$ were able to escape their initial vortex cells without becoming trapped anywhere. At large times, we observed that the MSD of all particles and the single-particle SD of individual particles became indistinguishable, both exhibiting a quadratic scaling with time. This behaviour signifies that all particles demonstrated ballistic motion, irrespective of variations in $\textrm{St}$, $\textrm{R}$, or their initial positions.  

%%%%%%%%%%%%%%%%%%%%%%%%%%%%%%%%%%%%%%%%%%%%%%%%%%%%%%%%%%%%%%%%%%%%%%%%%%%%%%
\begin{acknowledgments}
P.K. acknowledges insightful discussions with Divya Jaganathan on the integration of the history force, which contributed to the refinement of this work. A.V.S.N. thanks the Prime Minister’s Research Fellows (PMRF) scheme, Ministry of Education, Government of India.
\end{acknowledgments}



\end{document}